\documentclass[prl,aps,10pt,twocolumn,nofootinbib,preprintnumbers]{revtex4-1}
\usepackage{graphicx}
\usepackage{amsmath}
\usepackage{hyperref}
\usepackage{xspace}
\usepackage[utf8]{inputenc}
\usepackage{color}
\usepackage{todonotes}

\def\beq{\begin{equation}}
\def\beqn{\begin{eqnarray}}
\def\eeq{\end{equation}}
\def\eeqn{\end{eqnarray}}

\newcommand\GeV{\ensuremath{\mathrm{GeV}}\xspace}
\newcommand\TeV{\ensuremath{\mathrm{TeV}}\xspace}
\newcommand\ZZ{\ensuremath{\mathrm{ZZ}}\xspace}
\newcommand\WW{\ensuremath{\mathrm{W^{+}W^{-}}}\xspace}
\newcommand\mz{\ensuremath{m_{\mathrm{Z}}}\xspace}
\newcommand\pt{\ensuremath{p^{T}}\xspace}
\newcommand\mfl{\ensuremath{m_{{\mathrm4}\ell}}\xspace}
\newcommand\mll{\ensuremath{m_{\ell\ell}}\xspace}
\newcommand\kt{k_{\scriptscriptstyle\rm T}}

\newcommand\HWpp{{\tt HERWIG++}\xspace}
\newcommand\Herwigpp\HWpp
\newcommand\HERWIGPP\HWpp

\newcommand\PYTHIAn{{\tt PYTHIA \!\!8}\xspace}

\newcommand\POWHEG{{\tt POWHEG}}
\newcommand\POWHEGBOX{{\tt POWHEG BOX}\xspace}
\newcommand\POWHEGBOXVTWO{{\tt POWHEG BOX V2}\xspace}

\newcommand\FASTJET{{\tt Fastjet}\xspace}

\begin{document}

\title{ZZ production in gluon fusion at NLO matched to parton-shower}
\preprint{\vbox{\hbox{CERN-TH-2016-197}\hbox{TTP16-038}}}
\author{Simone Alioli}
\email{simone.alioli@cern.ch}
\affiliation{CERN Theory Division, CH-1211, Geneva 23, Switzerland}
\author{Fabrizio Caola}
\email{fabrizio.caola@cern.ch}
\affiliation{CERN Theory Division, CH-1211, Geneva 23, Switzerland}
\author{Gionata Luisoni}
\email{gionata.luisoni@cern.ch}
\affiliation{CERN Theory Division, CH-1211, Geneva 23, Switzerland}
\author{Raoul R\"ontsch}
\email{raoul.roentsch@kit.edu}
\affiliation{Institute for Theoretical Particle Physics, KIT, Karlsruhe, Germany}

\date{\today}

\begin{abstract}

We present a calculation of the next-to-leading order (NLO) QCD
corrections to the hadroproduction process $gg\to ZZ \to e^+e^- \mu^+
\mu^-$, matched to the parton shower in the \POWHEG{} framework.  We
take advantage of the \POWHEGBOX{} tool for the implementation and
rely on \PYTHIAn for the showering and hadronization stages.  We fully
include $\gamma^*/Z$ interference effects, while also covering the
single-resonant region. For this phenomenological study we focus on
four lepton production as a signal process, neglecting all quark mass
effects as well as the Higgs-mediated contributions, which are known
to be subdominant in this case. We provide predictions from our
simulations for the 13 TeV LHC Run II setup, including realistic
experimental cuts.
\end{abstract}

\keywords{QCD,NLO,Parton Shower}

\maketitle

\section{Introduction}

 During the Run I of the Large Hadron Collider (LHC), the production of
a pair of vector bosons was one of the processes studied in
greatest detail~\cite{Aad:2014wra,Aad:2015rka,CMS:2016kxu,Aad:2016wpd,Khachatryan:2016txa,Khachatryan:2015sga}.
The Higgs boson was initially discovered through its decay into dibosons,
and this decay channel continues to be important for measuring
the properties of the Higgs~\cite{Aad:2012tfa,Chatrchyan:2012xdj,Khachatryan:2015yvw,Aad:2015lha}.
First LHC Run II data at
$13$ TeV recently became available recently from both ATLAS~\cite{Aad:2015zqe} and CMS~\cite{Khachatryan:2016txa}.
Further LHC results with higher statistics will enable a program of
precision Higgs measurements, including its coupling to vector bosons
through its decay to \ZZ or \WW.
Additionally, diboson production is a benchmark process for precision tests of the Standard Model (SM),
while also providing constraints on anomalous gauge bosons couplings~\cite{Chatrchyan:2012sga,CMS:2014xja,Aad:2012awa}.
Without any direct sign of new physics, such indirect searches become increasingly relevant,
and the need for high precision becomes more important.

An essential requirement of such a program is the availability of
theoretical predictions for both signal and backgrounds which match
the experimental precision.  In the case of diboson, the lowest order
production mechanism is quark-antiquark annihilation.  The
next-to-next-to-leading order (NNLO) corrections to these processes
have been
computed~\cite{Grazzini:2016ctr,Gehrmann:2014fva,Grazzini:2016swo,Grazzini:2016hai,Grazzini:2015nwa,Grazzini:2013bna,Grazzini:2015hta,Cascioli:2014yka,Catani:2011qz,Campbell:2016yrh},
and have been matched to
resummation of the transverse
momentum of the diboson~\cite{Grazzini:2015wpa} and of the hardest jet~\cite{Dawson:2016ysj}.
Included in these
corrections is the contribution from gluon-initiated production, which
proceeds through a quark loop since the gluons do not couple directly
to electroweak gauge bosons.  Thus the leading order (LO) contribution
to the $gg$ channel is given by a one-loop amplitude and first enters
the overall diboson production rate at $\mathcal{O}(\alpha_s^2)$,
i.e. at NNLO.  It has been known for some time that these
contributions increase the cross section by approximately 5\%-15\%,
enhanced by the large gluonic flux~\cite{Binoth:2005ua,Binoth:2006mf,Binoth:2008pr}.
Furthermore,
since these contributions have LO-like scale uncertainty, they are
responsible for the majority of the residual scale uncertainty at
NNLO.

The NLO corrections for gluon-induced diboson production were recently
computed both for \ZZ~\cite{Caola:2015psa} and for
\WW~\cite{Caola:2015rqy}\footnote{The interference with the Higgs
  boson production channel was also recently computed in
  Refs.~\cite{Caola:2016trd,Campbell:2016ivq}.}.  There it was shown
that the NLO corrections further enhance the production rate,
resulting in an overall increase of the predictions for ZZ production
at the level of $\sim$5\%.  This exceeds the scale variation
uncertainty of the NNLO computation, making its inclusion important
for precision phenomenology.

In order to obtain accurate predictions with generic fiducial cuts,
the implementation of gluon-initiated diboson processes in a parton
shower framework is highly desirable.  In this paper, we present
results for the NLO QCD corrections to \ZZ production in the gluon
fusion channel, including the single-resonant region, matched to
parton shower within the \POWHEG{} framework. We do not consider
Higgs-mediated contributions and neglect quark-mass effects
throughout.  As such, our study represents a first step towards a
complete matching of NLO gluon-induced diboson production to parton
shower.

The remainder of the paper is organized as follows. First we present
the computational setup. Then we present our results, first applying a
generic set of cuts, and then using realistic experimental
cuts. Finally, we conclude and a give brief outlook.

\section{Computational setup}
\label{sec:computational_setup}
\begin{figure}[t!]
  \centering
  \includegraphics[width=0.49\textwidth]{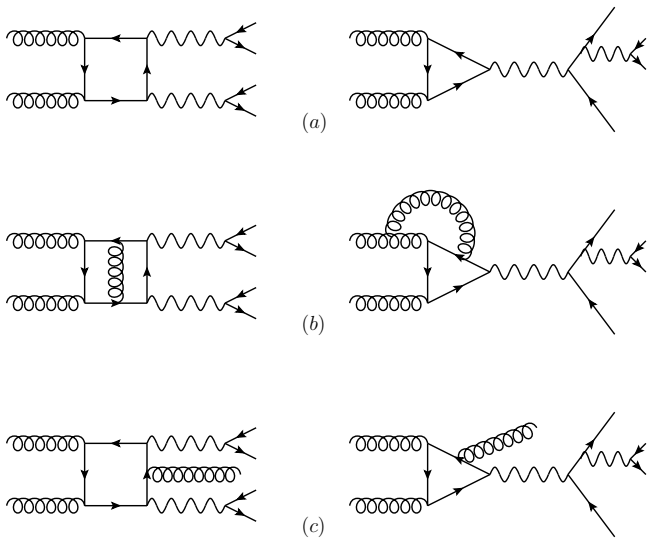}
  \caption{\label{fig:diagrams}%
    Example of Feynman diagrams considered at both leading \emph{(a)}
    and next-to-leading orders, for virtual \emph{(b)} and real
    \emph{(c)} QCD corrections.}
\end{figure}

In this study we focus on the $gg \to ZZ \to e^+e^-\mu^+\mu^-$ channel
as a signal process. 
We do not include Higgs-mediated contributions, 
which are mostly relevant around the Higgs peak and at very high invariant
masses (see e.g. Refs.~\cite{Kauer:2012hd,Campbell:2016ivq}
and~\cite{Caola:2016trd}).
The fixed-order computation is performed using the same strategy presented in
reference~\cite{Caola:2015psa}, which we recapitulate here briefly.

We consider \ZZ production via gluon fusion through a loop of massless quarks only.
At leading order
there are two types of loop-induced diagrams: boxes (an example is
shown on the left in Figure~\ref{fig:diagrams}-a) and triangles (right
in Figure~\ref{fig:diagrams}-a). The contribution from the latter
cancels within a massless quark family at all orders.  Since we are
considering five light flavors and neglecting contributions from the
massive top quark, for which the two-loop diagrams are not known, the
cancellation of these contributions between top and bottom quarks is
broken, giving rise to an anomaly. In order to avoid this we neglect
triangle diagrams at all orders, and work with five active massless
flavors, see Ref.~\cite{Caola:2015psa} for further details. At LO, the
neglected contributions affect the total $gg$-initiated cross-section
at the level of $1\%$, and restrict the validity of the predictions to
invariant masses of the four lepton system, $\mfl$, smaller than
roughly twice the top-quark mass~\cite{Kauer:2012hd,MCFM}.  As a
consequence, in our study we only show results in this kinematic
region.

At NLO, we need two-loop amplitudes for $gg\to e^+e^-\mu^+\mu^-$ and
one-loop amplitudes with one extra gluon in the final state.
Representative Feynman diagrams are shown in
Figure~\ref{fig:diagrams}-b and ~\ref{fig:diagrams}-c,
respectively. The two-loop amplitudes were recently computed in
Refs.~\cite{Caola:2015ila,vonManteuffel:2015msa} for internal massless
quarks. For this study we used the implementation of these amplitudes in the
\texttt{ggvvamp C++} package ~\cite{ggvvamp}. For the real-emission
amplitudes, we implemented the result computed in
reference~\cite{Caola:2015psa}, which provides fast and stable
predictions, including the soft and collinear regions.
At variance with the results presented in Ref.~\cite{Caola:2015psa},
in this work we have also included the real radiation contributions of
the form depicted on the right in Figure~\ref{fig:diagrams}-c.
Diagrams of this kind provide the only contribution to single resonant
production, since the triangle diagrams shown on the right of
Figure~\ref{fig:diagrams}-b are neglected, as previously discussed.
Therefore, the inclusion of the aforementioned
Figure~\ref{fig:diagrams}-c diagrams allows us to extend our
predictions to the single resonant region around the $Z$ boson peak.

For the real radiation contributions, the massless quark
approximation we are working in holds only for \mbox{$p^T_{4\ell}\lesssim
m_{\rm top}$}.  We stress that the contributions  to
the total cross-section outside this region are small and furthermore that our
calculation only has LO accuracy for $p^T_{4\ell} \neq 0$. For more reliable predictions in the
high-$p_T$ tail, an approach based on matrix-element corrections,
beyond the merging of the $0-$ and the $1-$jet samples already
presented in~\cite{Cascioli:2013gfa}, should be more suitable.  Here,
we won't focus on this aspect and leave these developments for a
separate investigation.

\begin{figure}[t!]
  \centering
  \includegraphics[width=0.49\textwidth]{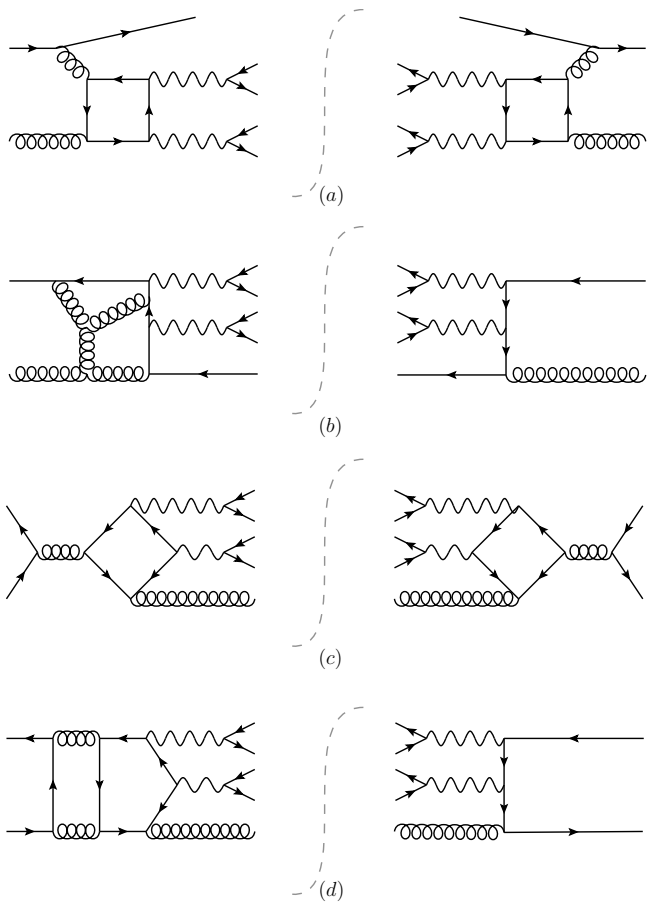}
  \caption{\label{fig:diagrams_n3lo}%
    Examples of Feynman diagrams contributing at NLO in the $qg$ and $q\bar{q}$-initiated channels. See text for comments.}
\end{figure}

Finally, we observe that formally the NLO corrections to $gg \to ZZ
\to e^+e^-\mu^+\mu^-$ include real-emission contributions of the
type $qg \to ZZ \to e^+e^-\mu^+\mu^- q$, $g q \to ZZ \to
e^+e^-\mu^+\mu^- q$ and $q \bar q \to ZZ \to e^+e^-\mu^+\mu^- g$. 
We note that to include quark-initiated channels in a complete
fashion, it is not sufficient to only consider the one-loop squared
contributions illustrated in Figures \mbox{\ref{fig:diagrams_n3lo}-a} and
\ref{fig:diagrams_n3lo}-c, but the full interferences contributing to
the N$^3$LO corrections to the quark-antiquark-initiated channel must
be taken into account. Examples of such diagrams are
illustrated in Figures~\ref{fig:diagrams_n3lo}-b and
~\ref{fig:diagrams_n3lo}-d. They require two-loop amplitudes which are
well beyond current technology.

Only including diagrams mediated
by a closed fermion loop, e.g. Figure~\ref{fig:diagrams_n3lo}-a, is possible, as
they are separately gauge invariant. However, we don't know any
reason of why these contributions should be dominant over the missing
ones, and sizable cancellations in the full result could in principle
take place.
As a consequence, in this study we work under the assumption that the
gluon luminosity is much larger than any quark parton distribution
function (PDF), and we omit all the contributions coming from diagrams
with quarks in the initial state.  This leads to a incomplete
compensation of factorization scale logarithms, parametrically
suppressed by the gluon/quark luminosity ratio, that should give an
indication of the size of the missing channels. It would be
interesting to study the effect of the inclusions of the
aforementioned loop-induced $qg$ channel. We leave this for a future
investigation.

We interfaced the NLO computation just described to the parton shower
using the \POWHEG{} method, implementing the \texttt{ggZZ} process
into the \POWHEGBOX{} program~\cite{Alioli:2010xd}\footnote{Our
  implementation heavily relies on new features available in the
  \POWHEGBOXVTWO{}, such as the parallelization of  the creation of integration grids and the parallel evaluation of the upper bounds for the
  event generation. Furthermore, the restriction to consider only the
  $gg$-initiated channel required a modification of the
  collinear-remnants contribution in the \POWHEGBOX{}, such that only
  the contributions coming from $g\to gg$ splittings were included.}.

Several checks of our implementation were performed against the code
used in~\cite{Caola:2015psa}, both at the level of the amplitude for
single phase space points, and at the level of integrated cross
sections with static and dynamical scales. Single phase space points
were also checked against
\textsc{GoSam}~\cite{Cullen:2011ac,Cullen:2014yla}, using
\textsc{Ninja}~\cite{vanDeurzen:2013saa,Peraro:2014cba}, and
\textsc{OpenLoops}~\cite{Cascioli:2011va}.  Furthermore, several checks
on the numerical stability were performed and a rescue system that
triggers the re-evaluation of the unstable phase-space points in
quadruple precision has been set up. Nonetheless, a technical cut
$\pt_{\mathrm{ZZ}} > 0.5$~\GeV has to be imposed to avoid
instabilities in the one-loop matrix elements. By varying the cut
value, we have checked that the neglected power-suppressed
contributions do not significantly change the total cross-section.

\section{Results}
\label{sec:results}
In this section we present results at LO, NLO and after interfacing
with the \PYTHIAn~\cite{Sjostrand:2007gs} parton shower.  We also include results at the
so-called Les Houches event (LHE) level, i.e.~after the first hard
emission generated with the \POWHEG{} method.

We consider center-of-mass energies of 8 and 13~\TeV, and consider two choices of the renormalization and factorization scales
\begin{align}
\mu=\mu_{R}=\mu_{F}&=m_{\mathrm{Z}}\,,
\end{align}
and
\begin{align}
\mu=\mu_{R}=\mu_{F}&=\frac{\mfl}{2}\,,
\end{align}
where
\begin{align}
\mfl^{2}=\left(p_{e^{+}}+p_{e^{-}}+p_{\mu^{+}}+p_{\mu^{-}}\right)^{2}\,.
\end{align}
  
For all the cases (both at LO and at NLO) we use the partonic luminosities and strong coupling from the \texttt{NNPDF30\_nlo\_as\_0118} set~\cite{Ball:2014uwa} and fix the
electroweak parameters to the following values:
\begin{align*}
m_{\mathrm{Z}}&=91.1876~\GeV\;; & \Gamma_{\mathrm{Z}}&=2.4952~\GeV\\
m_{\mathrm{W}}&=80.3980~\GeV\;; & \sin\theta_{w}&=0.2226\\
\alpha^{-1}&=132.3384
\end{align*}
Jets are reconstructed with the anti-$\kt$
algorithm~\cite{Cacciari:2008gp} as implemented in the \FASTJET{}
package~\cite{Cacciari:2005hq, Cacciari:2011ma}, with jet radius
$R=0.4$.
Furthermore, the following kinematical cuts are applied:
\begin{align}
5~\GeV&<\mll<180~\GeV,\label{eq:cut_mll}\\
60~\GeV&<\mfl<360~\GeV\label{eq:cut_m4l}\,.
\end{align}

\begin{table}[t!]
\centering
\small
\begin{tabular}{|c| c | c || c | c |}
  \hline
  [fb]
  &\multicolumn{2}{|c||}{$\mu=\mfl/2$}
  &\multicolumn{2}{|c|}{ $\mu=\mz$\phantom{\Big|}}     \\
  \hline
 CME
 &LO
 &NLO
 &LO
 &NLO\phantom{\Big|} \\
  \hline
  \phantom{1}8 \TeV\phantom{\Big|} & $1.60^{+0.41}_{-0.30}$ & $2.98^{+0.51}_{-0.41}$ & $1.62^{+0.42}_{-0.31}$ & $2.98^{+0.29}_{-0.40}$  \\
  13 \TeV          \phantom{\Big|} & $3.85^{+0.97}_{-0.70}$ & $6.98^{+1.14}_{-0.94}$ & $3.94^{+0.98}_{-0.71}$ & $7.22^{+1.04}_{-1.04}$  \\
  \hline
\end{tabular}
\caption{$gg \to ZZ\to e^{+} e^{-} \mu^{+} \mu^{-}$ total NLO cross sections and theoretical uncertainties stemming from scale variations
  for 8 and 13~\TeV. Results for both a fixed and a dynamical
  choice of renormalization and factorization scales are shown.}
\label{table:ggzz_total_XS}
\end{table}

\begin{figure}[t!]
  \centering
  \includegraphics[width=0.49\textwidth]{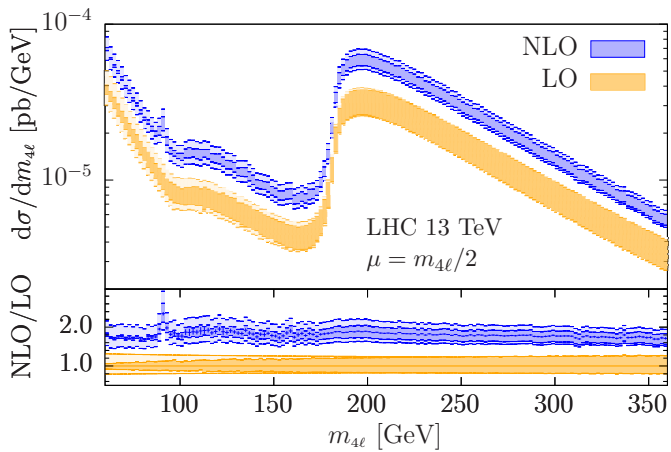}
  \caption{\label{fig:13Tev_mfl_lonlo}%
    Four-lepton invariant mass $\mfl$ distribution at LO and NLO for
    the LHC at 13~\TeV.  }
\end{figure}

\subsection{Fixed order}

In Table~\ref{table:ggzz_total_XS} we summarize the total cross
sections obtained for the setup just presented. The theoretical errors
reported are due to scale variations only. We estimate them by
independently varying the renormalization and factorization scales by
a factor of two around the reference value and excluding the two
extreme values of their ratio. We then use the minimum and maximum
values from the resulting seven-scale combination to assign the
uncertainties.

As already observed in~\cite{Caola:2015psa} the NLO corrections are
quite large and they lead to a stabilization of the scale uncertainty,
whose fractional value is roughly $15\%$ at both 8~\TeV and 13~\TeV.
These fractional uncertainties are slightly larger than those
previously reported in Ref.~\cite{Caola:2015psa}. We have verified
that the origin of the mismatch is due do the additional scale
combinations included in the envelope of our results and that perfect
agreement with the previous values is obtained when we only consider
$3-$point variations with equal renormalization and factorization
scales, as done by the authors of Ref.~\cite{Caola:2015psa}.

As expected, the two central scale values chosen ($M_Z$ and $\mfl/2$)
give very similar results. In the following, we use $\mfl/2$ as our
default, which is more suited over a wider range of invariant masses.

\begin{figure}[t!]
  \centering
  \includegraphics[width=0.49\textwidth]{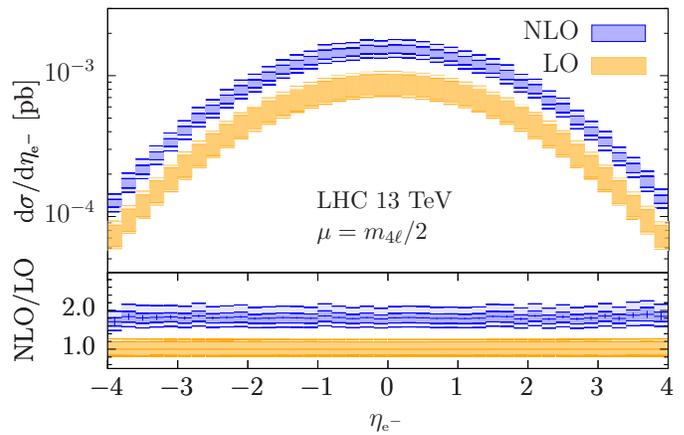}
  \caption{\label{fig:13Tev_etaem_lonlo}%
    Electron pseudorapidity distribution at LO and NLO for the LHC at
    13~\TeV.  }
\end{figure}

We now turn our attention to more differential observables, presenting
results for the LHC at 13~\TeV. We start by comparing the LO and NLO
curves of the four-lepton invariant mass distribution $\mfl$ in
Figure~\ref{fig:13Tev_mfl_lonlo}. Together with the main distribution,
we show the differential $K$-factor of the NLO predictions divided by
the LO ones in the lower inset. The lighter (darker) bands represent
the $7-$point ($3-$point) scale variation uncertainty, and in the
ratio plot we also display the statistical uncertainty in form of an
error bar. Both the LO and the NLO curves feature the typical enhancements
due to the photon propagator contribution at low values of $\mfl$, and
the steep increase at $\mfl\approx 180$~\GeV due to the $\ZZ$ double
resonant contribution. At NLO the single resonant channel opens up
leading to the peak at $\mfl\approx 90$~\GeV. Over the rest of the
spectrum the differential $K$-factor stays roughly constant at around
$1.8$.  This results in a flat $K$-factor for several inclusive
distributions.  As an example, we show in
Figure~\ref{fig:13Tev_etaem_lonlo} the pseudorapidity of the
electron.

\begin{figure}[t!]
  \centering
  \includegraphics[width=0.49\textwidth]{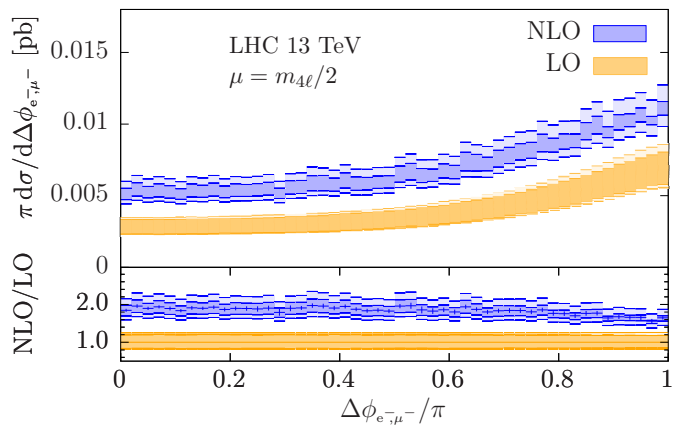}
  \caption{\label{fig:13Tev_dphiemmu_lonlo}%
    Azimuthal separation between electron and muon at LO and NLO for the LHC at
    13~\TeV. }
\end{figure}

Next, in Figure~\ref{fig:13Tev_dphiemmu_lonlo} we investigate the azimuthal separation between $e^-$ and
$\mu^-$, which provides interesting information about the diboson production mechanism~\cite{Bolognesi:2012mm}
A constant $K$-factor is again observed across almost the whole spectrum, with small deviations only visible in the region around $180$ degrees.

The situation is different for observables which are sensitive
to extra QCD radiation. 
\begin{figure}[t!]
  \centering
  \includegraphics[width=0.49\textwidth]{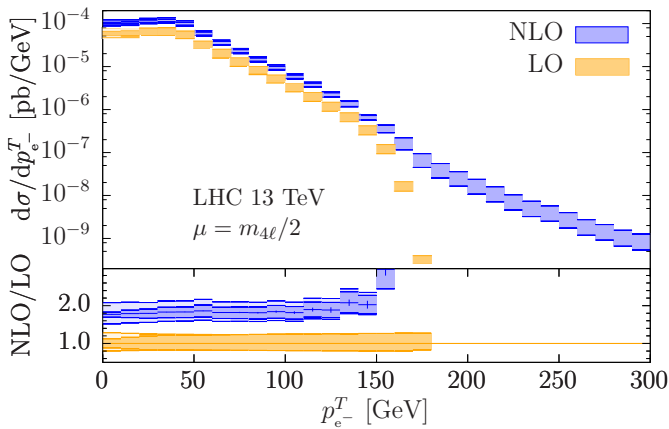}
  \caption{\label{fig:13Tev_ptem_lonlo}%
    Electron transverse momentum distribution at LO and NLO for the LHC at
    13~\TeV.  }
\end{figure}
An interesting example is the transverse momentum distribution of the electron,
shown in Figure~\ref{fig:13Tev_ptem_lonlo}.
At leading order the curve has an upper kinematical bound at
$p^{T}_{\mathrm{e^{-}}}=180$~\GeV due to the upper limit on $\mfl$
introduced in eq.~(\ref{eq:cut_m4l}). In the low end of the spectrum
this observable is predicted at NLO accuracy and it shows a
flat $K-$factor.  The additional radiation from the real emission
contribution allows the electrons to be produced with a transverse
momentum larger than the kinematic limit of $180$~\GeV. This means that, above
this value, the NLO curve effectively becomes LO. This is reflected in
the population of this region and in the enlarged scale uncertainty band.

A similar sensitivity to a kinematic threshold is also present
in the differential distribution of
$H^{T}_{\mathrm{Tot}}$ shown in
Figure~\ref{fig:13Tev_httot_lonlo} and defined as:
\begin{equation}
\label{eq:httot}
H^{T}_{\mathrm{Tot}}=\pt_{e^{+}}+\pt_{e^{-}}+\pt_{\mu^{+}}+\pt_{\mu^{-}}+\sum_{j\in\mathcal{J}}\pt_{j}\,,
\end{equation}
where the sum runs over the set $\mathcal{J}$ of final state
jets. At fixed NLO there can of course be at most one resolved jet,
due to the real radiation emission. The situation can however be more
involved after the shower, and we will comment further on this in the
next section.

\begin{figure}[t!]
  \centering
  \includegraphics[width=0.49\textwidth]{./plots/%
    LO_NLO_httot.pdf}
  \caption{\label{fig:13Tev_httot_lonlo}%
    $H^{T}_\mathrm{Tot}$ distribution at LO and NLO for the LHC at
    13~\TeV.}
\end{figure}

\subsection{Fixed order vs. \POWHEG{} first emission}

As a next step, we compare predictions at the NLO and at the Les
Houches event (LHE) level, meaning with the addition of the first hard
emission generated according to the \POWHEG{} method.  The results at
the LHE level are unphysical, but the comparison with the fixed-order
results and, later on, with the fully-showered ones helps in assessing how
big are the effects due to the exponentiation intrinsic in the \POWHEG{}
method and in separating them from the pure showering.

 In order to
avoid an excessive enhancement of the high-transverse momentum tail of the
$\ZZ$-pair and of the hardest-jet, previously observed in
similar \POWHEG{} implementations of processes with large $K$-factors
and discussed at length in Refs.~\cite{Alioli:2008tz,Alioli:2009je},
we have chosen to limit the amount of real radiation that gets
exponentiated by the Sudakov factor by setting the \texttt{hdamp}~\cite{Alioli:2008tz,Alioli:2009je}
parameter in the \POWHEGBOX{} to $100$~\GeV. This effectively ensures that
we smoothly recover the exact NLO result above that scale.

\begin{figure}[t!]
  \centering
  \includegraphics[width=0.49\textwidth]{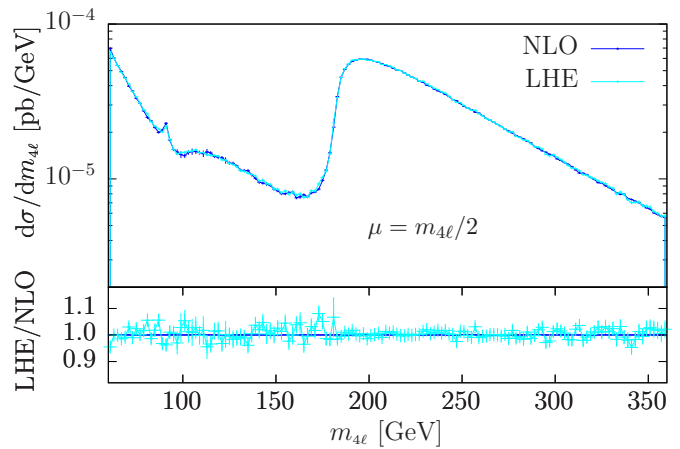}
  \caption{\label{fig:13Tev_m4l_nlolhe}%
    Invariant mass distribution of the four-lepton system at NLO and
    LHE-level at 13~\TeV.}
\end{figure}

Figure~\ref{fig:13Tev_m4l_nlolhe} shows again the four-lepton
invariant mass spectrum at NLO and LHE level. Apart from some
statistical fluctuations caused by the narrow binning, which we kept in
order to highlight the single-resonant peak at $\mfl=m_{\mathrm{Z}}$,
the agreement between LHE and NLO predictions is good over the whole
kinematical range. This is the expected result for observables which
are inclusive over the extra radiation generated by \POWHEG{}.
We have verified that similar results are obtained
for several other inclusive observables, e.g. the rapidities of the
leptons, or of the reconstructed $Z$-bosons.

\begin{figure}[t!]
  \centering
  \includegraphics[width=0.49\textwidth]{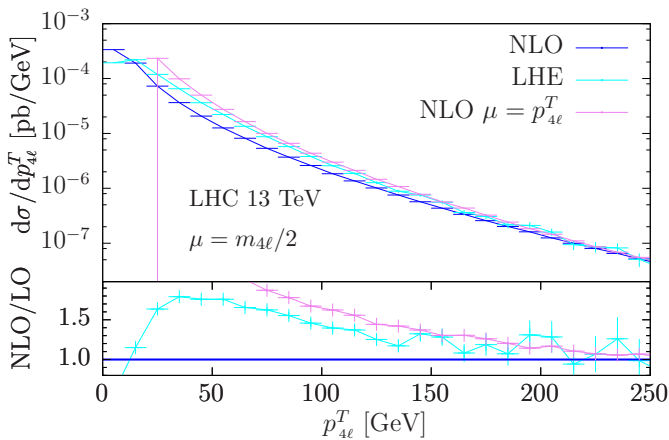}
  \caption{\label{fig:13Tev_ptZZ_nlolhe}%
    Transverse momentum distribution of the four-lepton system at NLO
    and LHE-level at 13~\TeV.}
\end{figure}

The situation is different when we consider the transverse momentum of
the four-lepton system, shown in
Figure~\ref{fig:13Tev_ptZZ_nlolhe}. Since the four-lepton system recoils
against the real radiation emission and has vanishing transverse
momentum when the emission becomes soft or collinear, this observable
is directly sensitive to the real radiation. The NLO curve diverges for
$p^{T}_{\mathrm{4\ell}}\to0$. However, when the real radiation is
weighted by the Sudakov form factor in the LHE-level predictions, we
observe the effect of the Sudakov suppression and the distribution
becomes finite for vanishing transverse momenta. Far away from the
Sudakov region, for transverse momenta $p^{T}_{\mathrm{4\ell}}>150~\GeV$
the NLO and LHE gets closer, as expected following the usage of
the \texttt{hdamp} factor in the \POWHEG{} implementation.

We note, however, that even after the inclusion of
the \texttt{hdamp} factor an exact agreement between the NLO and LHE
results in the tail of the distribution should not be expected. This
is a consequence of the different choices for the renormalization and
factorization scales used in the two calculations. The matrix element
for the real radiation is indeed evaluated according to the \POWHEG{}
method at $\mu_{R}=\mu_{F}=p^T_{4\ell}$ for the LHE results, while for the
NLO results they are evaluated at $\mu_{R}=\mu_{F}=\mfl/2$.  In order to
quantify the effects of this discrepancy in
Figure~\ref{fig:13Tev_ptZZ_nlolhe} we also plot the fixed-order results
above $p^T_{4\l}>20$~GeV choosing $\mu_{R}=\mu_{F}=p^T_{4\ell}$. We see
that a reasonable agreement between the three curves is reached above
$200-250$~GeV, before the two NLO curves start to depart for higher
values of the scales (not shown in the plot).  In any case, we would like to stress that due
to the massless-quark approximation we are working in, the predictions
for $p^T_{4\ell}$ (or correspondingly $p^T_{j}$) should not be 
trusted for larger values of the transverse momentum, because the
effects of the massive top-quarks in the loop can no longer be
neglected.

\begin{figure}[t!]
  \centering 
  \includegraphics[width=0.49\textwidth]{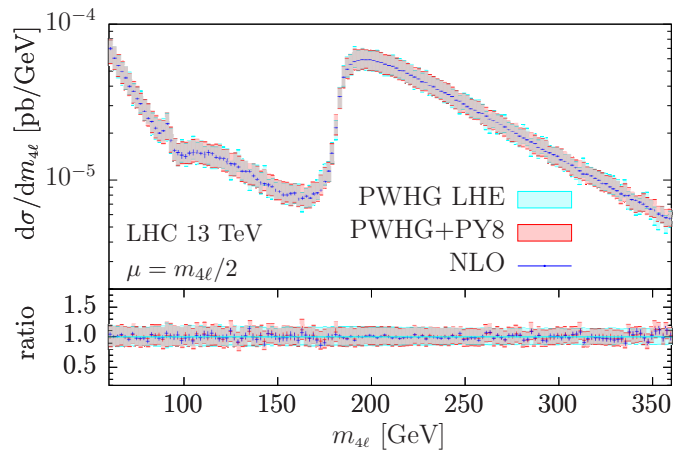}
  \caption{\label{fig:13Tev_m4l_lhepy8}%
    Invariant mass distribution of the four lepton system at the LHE
    level and after shower and hadronization with \PYTHIAn{}, compared
    to the fixed NLO curve.}
\end{figure}

\subsection{Showered results}

We now turn to the study of the impact of the parton shower. The
results showed in the following are produced using \PYTHIAn{} for the
showering and hadronization stages. In order to keep the analysis
simpler and to have a more direct comparison with theoretical
predictions at the partonic level, we have decided not to include
multiple parton interactions in the following plots.

We also remark that the limitation to only consider the
gluon-initiated channel that is used at the fixed-order or LHE level
is removed when we interface with the parton shower, which is
free to generate $q \to q g $ initial-state splittings.  This is
allowed by the unitary of the backward-evolved parton shower, which
for a given hard process produces the same total cross-section
irrespective of the partonic splittings allowed. To quantify the
impact of the inclusion of the quarks in the shower, we have also
studied the extreme case where the shower is only allowed to perform
$g\to gg$ splittings\footnote{This can be achieved by setting
  \texttt{SpaceShower:nQuarkIn = 0 } in \PYTHIAn{}. Note that this
  removes quarks altogether, which is different from our large gluon
  flux approximation. As such, this only provides an upper bound on
  effects due to the presence of quarks in the shower.}. No
appreciable differences for differential distributions are found, apart from two expected exceptions.
First, the transverse momentum of the
hardest jet at very low values, which is clearly affected by the
number and type of splittings included in the Sudakov exponent. Second,
the inclusion of quarks leads to mildly harder transverse-momentum spectra, as already observed in~\cite{Cascioli:2013gfa}.

\begin{figure}[t!]
  \centering
  \includegraphics[width=0.49\textwidth]{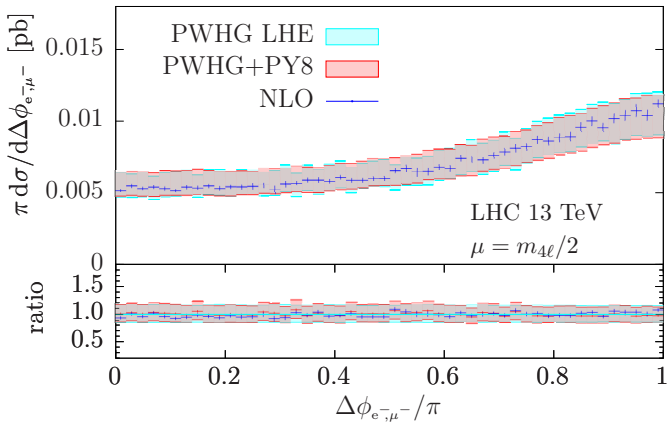}
  \caption{\label{fig:13Tev_dphiemmu_lhepy8}%
    Same as Figure~\ref{fig:13Tev_m4l_lhepy8} but for the
    azimuthal separation between the electron and the muon. }
\end{figure}

\begin{figure}[t!]
  \centering
  \includegraphics[width=0.49\textwidth]{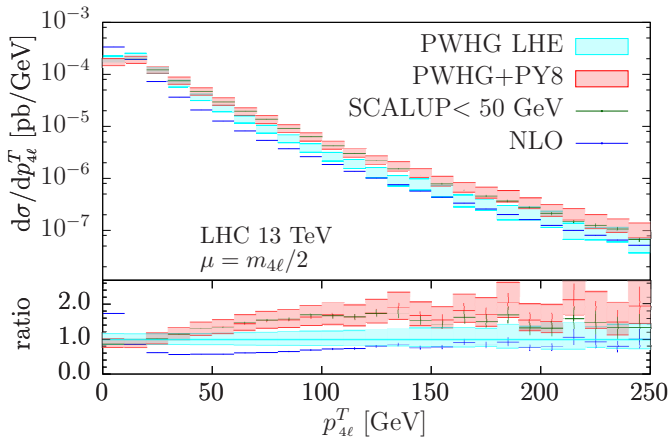}
  \caption{\label{fig:13Tev_pZZ_lhepy8}%
    Same as Figure~\ref{fig:13Tev_m4l_lhepy8} but for the transverse
    momentum of the four-lepton system.}
\end{figure}

\begin{figure}[t!]
  \centering
  \includegraphics[width=0.49\textwidth]{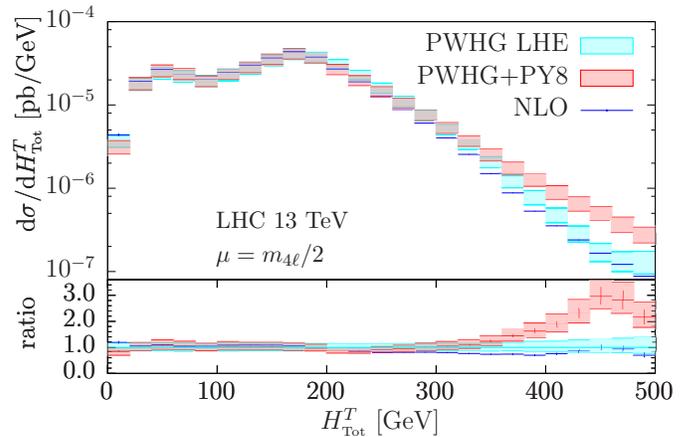}
  \caption{\label{fig:13Tev_httot_lhepy8}%
    Same as Figure~\ref{fig:13Tev_m4l_lhepy8} but for  $H^{T}_\mathrm{Tot}$, with jets  reconstructed by the anti-$k_T$ algorithm with $R=0.4$.}
\end{figure}

\begin{figure}[t!]
  \centering
  \includegraphics[width=0.49\textwidth]{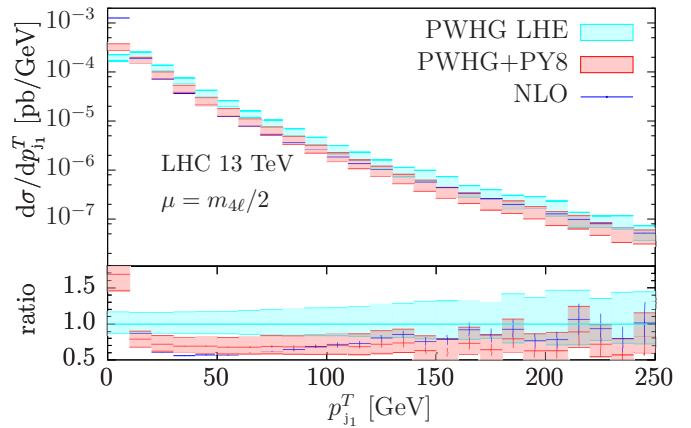}
  \caption{\label{fig:13Tev_ptJ_lhepy8}%
    Same as Figure~\ref{fig:13Tev_m4l_lhepy8} but for the transverse
    momentum of the hardest jet, reconstructed by the anti-$k_T$ algorithm with $R=0.4$.}
\end{figure}

\begin{figure}[t!]
  \centering
  \includegraphics[width=0.49\textwidth]{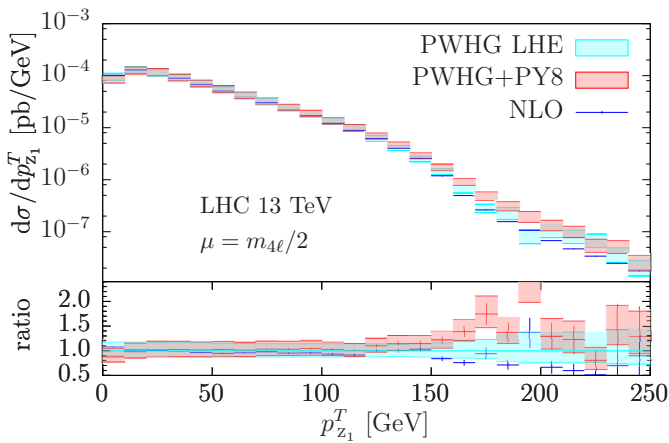}
  \caption{\label{fig:13Tev_ptZ_lhepy8}%
    Same as Figure~\ref{fig:13Tev_m4l_lhepy8} but for the transverse
    momentum of the $Z$ boson whose invariant mass is closer to the
    mass peak.}
\end{figure}

In Figures~\ref{fig:13Tev_m4l_lhepy8}-\ref{fig:13Tev_ptZ_lhepy8} we
compare the showered results to the NLO and LHE results at the nominal scale
$\mu=\mfl/2$. In all the observables we note a scale uncertainty which
varies around $20\%$, as is the case for fixed-order predictions.

For observables which are inclusive over the extra radiation, we note
an excellent agreement between the LHE-level results and the
\POWHEG{}+\PYTHIAn{} predictions. This is also true for the theory
uncertainty bands which overlap almost perfectly. As one would expect,
the parton shower does not have a strong influence on these
quantities. This is shown for the four-lepton invariant mass
distribution, in Figure~\ref{fig:13Tev_m4l_lhepy8}, in which the
single-resonant peak is still clearly visible.  We have verified that
the shower has a similarly small effect for several other inclusive
distributions, including the rapidities of the leptons and of the
$Z$-bosons. As a further example, we show the azimuthal separation
between $e^-$ and $\mu^-$ in Figure~\ref{fig:13Tev_dphiemmu_lhepy8}.

By contrast, the parton shower has a larger impact on the transverse
momentum of the four lepton system,
Figure~\ref{fig:13Tev_pZZ_lhepy8}. At small transverse momenta it
undershoots the LHE-level prediction by roughly $10\%$, but above
30~\GeV it becomes larger than the LHE-level results, reaching a
plateau around 150~\GeV, where the ratio between the two predictions
is between 1.5 and 2.  The large discrepancy between the showered
results and the fixed order (or the LHE-level) ones in the tail of the
distribution, which are still however roughly compatible given the
correspondingly large LO scale variations, can be explained by the
fact that by adding further radiation the shower increases the
transverse momentum of the color-neutral four lepton system, which has
to recoil against the sum of all emitted particles.  This can be
further demonstrated by lowering the starting scale for the \PYTHIAn{}
showering: for example, in Figure~\ref{fig:13Tev_pZZ_lhepy8} we also
include predictions where we have limited the hardness of shower
emissions to be lower than $50$~GeV, irrespective of the hardness of
the first \POWHEG{} emission\footnote{This is done by limiting the
  \texttt{SCALUP} value to $50$~GeV.}. The resulting predictions in
the large $p^T_{4\ell}$ region are closer to the NLO curve, due to the
reduced \PYTHIAn{} activity.  The same effect is seen in
Figure~\ref{fig:13Tev_httot_lhepy8} for the scalar sum of the
transverse momenta defined in Eq.~\ref{eq:httot}.

On the contrary, a similar enhancement is not expected when looking at
the transverse momentum of the hardest jet in the event.  Indeed, the
shower emissions are by construction subdominant with respect to the
leading jet and on average are separated enough not to be clustered
with it. Therefore, while the shower has a larger effect on the
transverse momentum of the colorless recoiling system, it should not
significantly affect the leading-jet spectrum. This is observed in
Figure \ref{fig:13Tev_ptJ_lhepy8}, which only displays a mild
softening of the leading jet $p_T$ with respect to the LHE results,
due to radiation off the jet.

In our analysis the two $Z$-bosons are reconstructed according to their
invariant mass. Event by event we distinguish two $Z$-bosons: the one
whose invariant mass is closer to $m_{\mathrm{Z}}$, labeled $Z_1$, and
the one further away, labeled $Z_2$. Since we consider the
$e^+e^-\mu^+\mu^-$ final state, $Z_1$ and $Z_2$ are always
reconstructed by opposite sign leptons from the same family. However, this
procedure allows to uniquely define the two $Z$ bosons also
for final states with equal pairs of leptons.

The transverse momentum
of the $Z_1$ boson, displayed in Figure~\ref{fig:13Tev_ptZ_lhepy8}, is almost
unaffected by parton shower corrections for values of the transverse
momentum smaller than 150~\GeV. For harder values of $p^{T}_{Z_1}$ the
shower increases the cross section. This effect is related to the
crossing of the kinematic threshold already observed in the comparison
of LO and NLO predictions for the electron transverse momentum in
Figure~\ref{fig:13Tev_ptem_lonlo}.

\subsection{ATLAS fiducial cuts}
\label{sec:atlas_cuts}
Before concluding we present results obtained applying fiducial cuts
similar to the ones used by the ATLAS collaboration
in~\cite{ATLAS:2013gma}, namely:

\begin{align}
80~\GeV &< \mfl < 350~\GeV,\,\nonumber\\
66~\GeV &< \mll < 160~\GeV,\,\nonumber\\
\Delta R_{\ell\ell}  &> 0.2,\,\\
\pt_{\ell}  &> 7~\GeV,\,\nonumber\\
|\eta_{\ell}| &> 2.7.\nonumber
\end{align}

Within these fiducial cuts, the resulting NLO cross section at 13~\TeV
is
\begin{equation}
\sigma^{\rm{fid.}} = 4.57^{+0.71}_{-0.59}\ \rm{fb}.
\end{equation}

\begin{figure}[t!]
  \centering
  \includegraphics[width=0.49\textwidth]{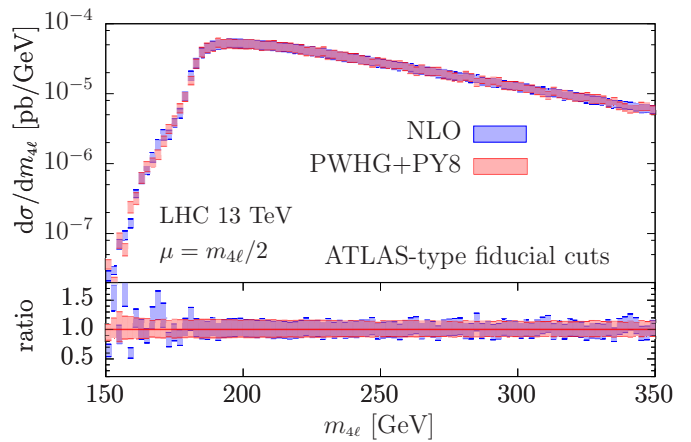}
  \caption{\label{fig:13Tev_m4l_atlas}%
    Invariant mass distribution of the four-lepton system at NLO and after shower
    and hadronization with \PYTHIAn{} when ATLAS fiducial cuts are
    applied.}
\end{figure}
\begin{figure}[t!]
  \centering
  \includegraphics[width=0.49\textwidth]{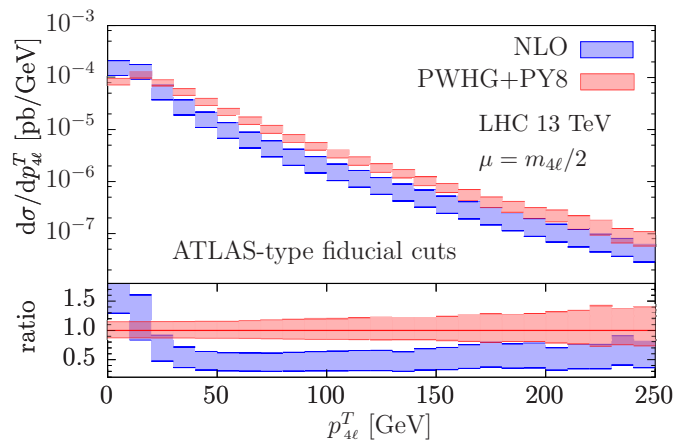}
  \caption{\label{fig:13Tev_ptZZ_atlas}%
    Transverse momentum distribution of the four-lepton system at NLO and after
    shower and hadronization with \PYTHIAn{} when ATLAS fiducial cuts
    are applied.}
\end{figure}
\begin{figure}[t!]
  \centering
  \includegraphics[width=0.49\textwidth]{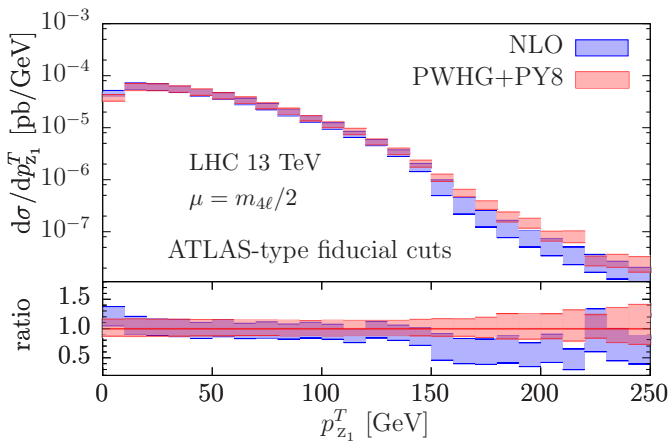}
  \caption{\label{fig:13Tev_ptZ_atlas}%
    Transverse momentum distribution of the $Z$ boson, whose invariant
    mass is closer to the mass peak, at NLO and after shower and hadronization
    with \PYTHIAn{} when ATLAS fiducial cuts are applied.}
\end{figure}
\begin{figure}[t!]
  \centering
  \includegraphics[width=0.49\textwidth]{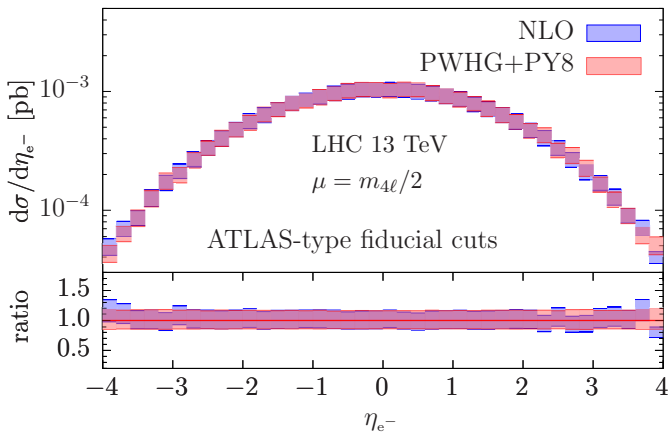}
  \caption{\label{fig:13Tev_etaem_atlas}%
    Rapidity distribution of the electron  at NLO and after shower and
    hadronization with \PYTHIAn{} when ATLAS fiducial cuts are applied.}
\end{figure}
\begin{figure}[t!]
  \centering
  \includegraphics[width=0.49\textwidth]{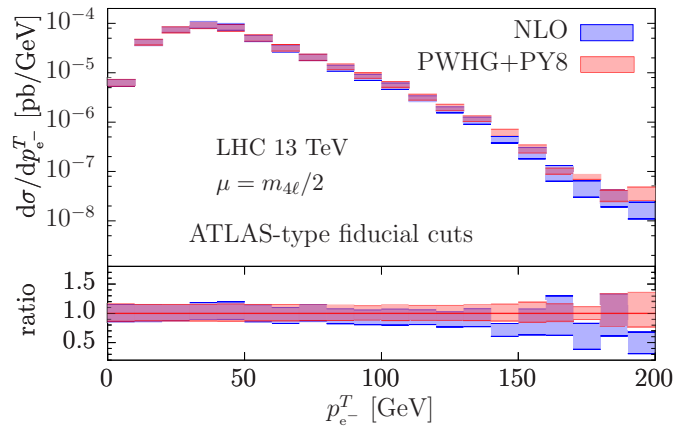}
  \caption{\label{fig:13Tev_ptem_atlas}%
    Transverse momentum distribution of the electron  at NLO and after shower and
    hadronization with \PYTHIAn{} when ATLAS fiducial cuts are applied.}
\end{figure}

In Figures~\ref{fig:13Tev_m4l_atlas}-\ref{fig:13Tev_ptem_atlas} we compare the corresponding
\POWHEG{}+\PYTHIAn{} predictions with pure NLO ones for the four-lepton invariant mass
and transverse momentum, for the transverse momentum of the $Z_1$
boson and for the pseudorapidity and transverse momentum of the
electron, respectively.  In the lower inset of each plot we show the
scale uncertainty band obtained with a 7-point variation of
renormalization and factorization scales, as explained in the previous
section.  The same features observed for the more inclusive analysis
are present in this fiducial region, as expected for such inclusive cuts.

\section{Conclusions and Outlook}
\label{sec:conclusions_outlook}
The production of a pair of $Z$ bosons plays a key role at the LHC,
not only as a further important test of the Standard Model, but in
relation to Higgs studies.  In this paper we have interfaced a NLO
computation for $\ZZ$-production in gluon fusion to a parton shower,
using the \POWHEGBOX{} framework.  The calculation has been performed
in the limit of the gluon PDF being much larger than any quark one. We
also neglected quark mass effects in the loops throughout.  In this
study we have primarily concentrated on the $\ZZ$-production process
as a signal and have consequently not attempted to include either the
Higgs-boson mediated channel or the interference between the two
production modes.  These effects are known to be important and
severely affect the production rates mostly around the
Higgs-boson resonant region and for large values of
$\mfl$~\cite{Kauer:2012hd,Campbell:2016ivq,Caola:2016trd}. The proper inclusion of all
the aforementioned effects in a NLO calculation matched to parton
shower will the subject of a separate investigation.

After interfacing with the parton shower in \PYTHIAn, our findings are
in agreement with the expectations. In particular, we observe that
quantities which are inclusive over the extra radiation do not receive
appreciable modifications by the showering stage. On the other hand,
there is a substantial effect due to the parton shower for quantities
that are more sensitive to the hadronic activity, even when the
observables are built exclusively using the four momenta of the
leptons coming from the $Z$ decays. A typical example is the
transverse momentum of the four-lepton system, which at NLO is
constrained by momentum conservation to recoil against the emitted
parton. Other situations where the parton shower provides large
corrections appear in the presence of multiple shower emissions, which
allow observables to evade kinematical bounds that would be otherwise present at
fixed-order.  This is e.g. observed in the transverse momentum of
hardest $Z$-boson, above the kinematical bound set by the generation
cuts on $\mfl$.
We have found similar effects also when applying realistic experimental
cuts, modeled on those used by the ATLAS collaboration in a previous
analysis of the $\ZZ$ four-lepton final state. We provided predictions
 in this fiducial region for the 13 TeV LHC Run II.

There are several interesting further developments we didn't
investigate in this first study: for example the inclusion of
quark-mass effects in the loops as well as a detailed study of the
Higgs-mediated contributions and their relevance for off-shell Higgs
analyses. Also, it would be interesting to explore the impact of the
matching to different parton showers and their comparison with
matrix-elements corrected approaches. We leave these for future
investigations.

\section{Acknowledgments}
We thank J.~Lindert for help and support in comparing the analytic
amplitudes used in this calculation with \textsc{OpenLoops}.  We also
thank E.~Re and P.~Nason for useful conversations and suggestions
about the optimal usage of new features of the \POWHEGBOXVTWO{}.

The numerical computations were performed at the RZG -- the
Rechenzentrum Garching near Munich Garching.  This work was supported
by the COFUND Fellowship under grant agreement PCOFUND-GA-2012-600377
(SA), and by the German Federal Ministry for Education and Research (BMBF) under grant 05H15VKCCA (RR).

\section*{References}
\bibliography{paper}

\begin{thebibliography}{54}%
\makeatletter
\providecommand \@ifxundefined [1]{%
 \@ifx{#1\undefined}
}%
\providecommand \@ifnum [1]{%
 \ifnum #1\expandafter \@firstoftwo
 \else \expandafter \@secondoftwo
 \fi
}%
\providecommand \@ifx [1]{%
 \ifx #1\expandafter \@firstoftwo
 \else \expandafter \@secondoftwo
 \fi
}%
\providecommand \natexlab [1]{#1}%
\providecommand \enquote  [1]{``#1''}%
\providecommand \bibnamefont  [1]{#1}%
\providecommand \bibfnamefont [1]{#1}%
\providecommand \citenamefont [1]{#1}%
\providecommand \href@noop [0]{\@secondoftwo}%
\providecommand \href [0]{\begingroup \@sanitize@url \@href}%
\providecommand \@href[1]{\@@startlink{#1}\@@href}%
\providecommand \@@href[1]{\endgroup#1\@@endlink}%
\providecommand \@sanitize@url [0]{\catcode `\\12\catcode `\$12\catcode
  `\&12\catcode `\#12\catcode `\^12\catcode `\_12\catcode `\%12\relax}%
\providecommand \@@startlink[1]{}%
\providecommand \@@endlink[0]{}%
\providecommand \url  [0]{\begingroup\@sanitize@url \@url }%
\providecommand \@url [1]{\endgroup\@href {#1}{\urlprefix }}%
\providecommand \urlprefix  [0]{URL }%
\providecommand \Eprint [0]{\href }%
\providecommand \doibase [0]{http://dx.doi.org/}%
\providecommand \selectlanguage [0]{\@gobble}%
\providecommand \bibinfo  [0]{\@secondoftwo}%
\providecommand \bibfield  [0]{\@secondoftwo}%
\providecommand \translation [1]{[#1]}%
\providecommand \BibitemOpen [0]{}%
\providecommand \bibitemStop [0]{}%
\providecommand \bibitemNoStop [0]{.\EOS\space}%
\providecommand \EOS [0]{\spacefactor3000\relax}%
\providecommand \BibitemShut  [1]{\csname bibitem#1\endcsname}%
\let\auto@bib@innerbib\@empty
\bibitem [{\citenamefont {Aad}\ \emph {et~al.}(2014)\citenamefont {Aad} \emph
  {et~al.}}]{Aad:2014wra}%
  \BibitemOpen
  \bibfield  {author} {\bibinfo {author} {\bibfnamefont {G.}~\bibnamefont
  {Aad}} \emph {et~al.} (\bibinfo {collaboration} {ATLAS}),\ }\href {\doibase
  10.1103/PhysRevLett.112.231806} {\bibfield  {journal} {\bibinfo  {journal}
  {Phys. Rev. Lett.}\ }\textbf {\bibinfo {volume} {112}},\ \bibinfo {pages}
  {231806} (\bibinfo {year} {2014})},\ \Eprint {http://arxiv.org/abs/1403.5657}
  {arXiv:1403.5657 [hep-ex]} \BibitemShut {NoStop}%
\bibitem [{\citenamefont {Aad}\ \emph {et~al.}(2016{\natexlab{a}})\citenamefont
  {Aad} \emph {et~al.}}]{Aad:2015rka}%
  \BibitemOpen
  \bibfield  {author} {\bibinfo {author} {\bibfnamefont {G.}~\bibnamefont
  {Aad}} \emph {et~al.} (\bibinfo {collaboration} {ATLAS}),\ }\href {\doibase
  10.1016/j.physletb.2015.12.048} {\bibfield  {journal} {\bibinfo  {journal}
  {Phys. Lett.}\ }\textbf {\bibinfo {volume} {B753}},\ \bibinfo {pages} {552}
  (\bibinfo {year} {2016}{\natexlab{a}})},\ \Eprint
  {http://arxiv.org/abs/1509.07844} {arXiv:1509.07844 [hep-ex]} \BibitemShut
  {NoStop}%
\bibitem [{CMS(2016)}]{CMS:2016kxu}%
  \BibitemOpen
  \href@noop {} {\bibfield  {journal} {\bibinfo  {journal}
  {CMS-PAS-SMP-15-012}\ } (\bibinfo {year} {2016})}\BibitemShut {NoStop}%
\bibitem [{\citenamefont {Aad}\ \emph {et~al.}(2016{\natexlab{b}})\citenamefont
  {Aad} \emph {et~al.}}]{Aad:2016wpd}%
  \BibitemOpen
  \bibfield  {author} {\bibinfo {author} {\bibfnamefont {G.}~\bibnamefont
  {Aad}} \emph {et~al.} (\bibinfo {collaboration} {ATLAS}),\ }\href {\doibase
  10.1007/JHEP09(2016)029} {\bibfield  {journal} {\bibinfo  {journal} {JHEP}\
  }\textbf {\bibinfo {volume} {09}},\ \bibinfo {pages} {029} (\bibinfo {year}
  {2016}{\natexlab{b}})},\ \Eprint {http://arxiv.org/abs/1603.01702}
  {arXiv:1603.01702 [hep-ex]} \BibitemShut {NoStop}%
\bibitem [{\citenamefont {Khachatryan}\ \emph
  {et~al.}(2016{\natexlab{a}})\citenamefont {Khachatryan} \emph
  {et~al.}}]{Khachatryan:2016txa}%
  \BibitemOpen
  \bibfield  {author} {\bibinfo {author} {\bibfnamefont {V.}~\bibnamefont
  {Khachatryan}} \emph {et~al.} (\bibinfo {collaboration} {CMS}),\ }\href@noop
  {} {\bibfield  {journal} {\bibinfo  {journal} {Submitted to: Phys. Lett. B}\
  } (\bibinfo {year} {2016}{\natexlab{a}})},\ \Eprint
  {http://arxiv.org/abs/1607.08834} {arXiv:1607.08834 [hep-ex]} \BibitemShut
  {NoStop}%
\bibitem [{\citenamefont {Khachatryan}\ \emph
  {et~al.}(2016{\natexlab{b}})\citenamefont {Khachatryan} \emph
  {et~al.}}]{Khachatryan:2015sga}%
  \BibitemOpen
  \bibfield  {author} {\bibinfo {author} {\bibfnamefont {V.}~\bibnamefont
  {Khachatryan}} \emph {et~al.} (\bibinfo {collaboration} {CMS}),\ }\href
  {\doibase 10.1140/epjc/s10052-016-4219-1} {\bibfield  {journal} {\bibinfo
  {journal} {Eur. Phys. J.}\ }\textbf {\bibinfo {volume} {C76}},\ \bibinfo
  {pages} {401} (\bibinfo {year} {2016}{\natexlab{b}})},\ \Eprint
  {http://arxiv.org/abs/1507.03268} {arXiv:1507.03268 [hep-ex]} \BibitemShut
  {NoStop}%
\bibitem [{\citenamefont {Aad}\ \emph {et~al.}(2012)\citenamefont {Aad} \emph
  {et~al.}}]{Aad:2012tfa}%
  \BibitemOpen
  \bibfield  {author} {\bibinfo {author} {\bibfnamefont {G.}~\bibnamefont
  {Aad}} \emph {et~al.} (\bibinfo {collaboration} {ATLAS Collaboration}),\
  }\href {\doibase 10.1016/j.physletb.2012.08.020} {\bibfield  {journal}
  {\bibinfo  {journal} {Phys.Lett.}\ }\textbf {\bibinfo {volume} {B716}},\
  \bibinfo {pages} {1} (\bibinfo {year} {2012})},\ \Eprint
  {http://arxiv.org/abs/1207.7214} {arXiv:1207.7214 [hep-ex]} \BibitemShut
  {NoStop}%
\bibitem [{\citenamefont {Chatrchyan}\ \emph {et~al.}(2012)\citenamefont
  {Chatrchyan} \emph {et~al.}}]{Chatrchyan:2012xdj}%
  \BibitemOpen
  \bibfield  {author} {\bibinfo {author} {\bibfnamefont {S.}~\bibnamefont
  {Chatrchyan}} \emph {et~al.} (\bibinfo {collaboration} {CMS}),\ }\href
  {\doibase 10.1016/j.physletb.2012.08.021} {\bibfield  {journal} {\bibinfo
  {journal} {Phys. Lett.}\ }\textbf {\bibinfo {volume} {B716}},\ \bibinfo
  {pages} {30} (\bibinfo {year} {2012})},\ \Eprint
  {http://arxiv.org/abs/1207.7235} {arXiv:1207.7235 [hep-ex]} \BibitemShut
  {NoStop}%
\bibitem [{\citenamefont {Khachatryan}\ \emph
  {et~al.}(2016{\natexlab{c}})\citenamefont {Khachatryan} \emph
  {et~al.}}]{Khachatryan:2015yvw}%
  \BibitemOpen
  \bibfield  {author} {\bibinfo {author} {\bibfnamefont {V.}~\bibnamefont
  {Khachatryan}} \emph {et~al.} (\bibinfo {collaboration} {CMS}),\ }\href
  {\doibase 10.1007/JHEP04(2016)005} {\bibfield  {journal} {\bibinfo  {journal}
  {JHEP}\ }\textbf {\bibinfo {volume} {04}},\ \bibinfo {pages} {005} (\bibinfo
  {year} {2016}{\natexlab{c}})},\ \Eprint {http://arxiv.org/abs/1512.08377}
  {arXiv:1512.08377 [hep-ex]} \BibitemShut {NoStop}%
\bibitem [{\citenamefont {Aad}\ \emph {et~al.}(2015)\citenamefont {Aad} \emph
  {et~al.}}]{Aad:2015lha}%
  \BibitemOpen
  \bibfield  {author} {\bibinfo {author} {\bibfnamefont {G.}~\bibnamefont
  {Aad}} \emph {et~al.} (\bibinfo {collaboration} {ATLAS}),\ }\href {\doibase
  10.1103/PhysRevLett.115.091801} {\bibfield  {journal} {\bibinfo  {journal}
  {Phys. Rev. Lett.}\ }\textbf {\bibinfo {volume} {115}},\ \bibinfo {pages}
  {091801} (\bibinfo {year} {2015})},\ \Eprint
  {http://arxiv.org/abs/1504.05833} {arXiv:1504.05833 [hep-ex]} \BibitemShut
  {NoStop}%
\bibitem [{\citenamefont {Aad}\ \emph {et~al.}(2016{\natexlab{c}})\citenamefont
  {Aad} \emph {et~al.}}]{Aad:2015zqe}%
  \BibitemOpen
  \bibfield  {author} {\bibinfo {author} {\bibfnamefont {G.}~\bibnamefont
  {Aad}} \emph {et~al.} (\bibinfo {collaboration} {ATLAS}),\ }\href {\doibase
  10.1103/PhysRevLett.116.101801} {\bibfield  {journal} {\bibinfo  {journal}
  {Phys. Rev. Lett.}\ }\textbf {\bibinfo {volume} {116}},\ \bibinfo {pages}
  {101801} (\bibinfo {year} {2016}{\natexlab{c}})},\ \Eprint
  {http://arxiv.org/abs/1512.05314} {arXiv:1512.05314 [hep-ex]} \BibitemShut
  {NoStop}%
\bibitem [{\citenamefont {Chatrchyan}\ \emph {et~al.}(2013)\citenamefont
  {Chatrchyan} \emph {et~al.}}]{Chatrchyan:2012sga}%
  \BibitemOpen
  \bibfield  {author} {\bibinfo {author} {\bibfnamefont {S.}~\bibnamefont
  {Chatrchyan}} \emph {et~al.} (\bibinfo {collaboration} {CMS}),\ }\href
  {\doibase 10.1007/JHEP01(2013)063} {\bibfield  {journal} {\bibinfo  {journal}
  {JHEP}\ }\textbf {\bibinfo {volume} {01}},\ \bibinfo {pages} {063} (\bibinfo
  {year} {2013})},\ \Eprint {http://arxiv.org/abs/1211.4890} {arXiv:1211.4890
  [hep-ex]} \BibitemShut {NoStop}%
\bibitem [{\citenamefont {Khachatryan}\ \emph {et~al.}(2015)\citenamefont
  {Khachatryan} \emph {et~al.}}]{CMS:2014xja}%
  \BibitemOpen
  \bibfield  {author} {\bibinfo {author} {\bibfnamefont {V.}~\bibnamefont
  {Khachatryan}} \emph {et~al.} (\bibinfo {collaboration} {CMS}),\ }\href
  {\doibase 10.1016/j.physletb.2016.04.010, 10.1016/j.physletb.2014.11.059}
  {\bibfield  {journal} {\bibinfo  {journal} {Phys. Lett.}\ }\textbf {\bibinfo
  {volume} {B740}},\ \bibinfo {pages} {250} (\bibinfo {year} {2015})},\
  \bibinfo {note} {[Erratum: Phys. Lett.B757,569(2016)]},\ \Eprint
  {http://arxiv.org/abs/1406.0113} {arXiv:1406.0113 [hep-ex]} \BibitemShut
  {NoStop}%
\bibitem [{\citenamefont {Aad}\ \emph {et~al.}(2013)\citenamefont {Aad} \emph
  {et~al.}}]{Aad:2012awa}%
  \BibitemOpen
  \bibfield  {author} {\bibinfo {author} {\bibfnamefont {G.}~\bibnamefont
  {Aad}} \emph {et~al.} (\bibinfo {collaboration} {ATLAS}),\ }\href {\doibase
  10.1007/JHEP03(2013)128} {\bibfield  {journal} {\bibinfo  {journal} {JHEP}\
  }\textbf {\bibinfo {volume} {03}},\ \bibinfo {pages} {128} (\bibinfo {year}
  {2013})},\ \Eprint {http://arxiv.org/abs/1211.6096} {arXiv:1211.6096
  [hep-ex]} \BibitemShut {NoStop}%
\bibitem [{\citenamefont {Grazzini}\ \emph
  {et~al.}(2016{\natexlab{a}})\citenamefont {Grazzini}, \citenamefont
  {Kallweit}, \citenamefont {Pozzorini}, \citenamefont {Rathlev},\ and\
  \citenamefont {Wiesemann}}]{Grazzini:2016ctr}%
  \BibitemOpen
  \bibfield  {author} {\bibinfo {author} {\bibfnamefont {M.}~\bibnamefont
  {Grazzini}}, \bibinfo {author} {\bibfnamefont {S.}~\bibnamefont {Kallweit}},
  \bibinfo {author} {\bibfnamefont {S.}~\bibnamefont {Pozzorini}}, \bibinfo
  {author} {\bibfnamefont {D.}~\bibnamefont {Rathlev}}, \ and\ \bibinfo
  {author} {\bibfnamefont {M.}~\bibnamefont {Wiesemann}},\ }\href {\doibase
  10.1007/JHEP08(2016)140} {\bibfield  {journal} {\bibinfo  {journal} {JHEP}\
  }\textbf {\bibinfo {volume} {08}},\ \bibinfo {pages} {140} (\bibinfo {year}
  {2016}{\natexlab{a}})},\ \Eprint {http://arxiv.org/abs/1605.02716}
  {arXiv:1605.02716 [hep-ph]} \BibitemShut {NoStop}%
\bibitem [{\citenamefont {Gehrmann}\ \emph {et~al.}(2014)\citenamefont
  {Gehrmann}, \citenamefont {Grazzini}, \citenamefont {Kallweit}, \citenamefont
  {Maierhöfer}, \citenamefont {von Manteuffel}, \citenamefont {Pozzorini},
  \citenamefont {Rathlev},\ and\ \citenamefont {Tancredi}}]{Gehrmann:2014fva}%
  \BibitemOpen
  \bibfield  {author} {\bibinfo {author} {\bibfnamefont {T.}~\bibnamefont
  {Gehrmann}}, \bibinfo {author} {\bibfnamefont {M.}~\bibnamefont {Grazzini}},
  \bibinfo {author} {\bibfnamefont {S.}~\bibnamefont {Kallweit}}, \bibinfo
  {author} {\bibfnamefont {P.}~\bibnamefont {Maierhöfer}}, \bibinfo {author}
  {\bibfnamefont {A.}~\bibnamefont {von Manteuffel}}, \bibinfo {author}
  {\bibfnamefont {S.}~\bibnamefont {Pozzorini}}, \bibinfo {author}
  {\bibfnamefont {D.}~\bibnamefont {Rathlev}}, \ and\ \bibinfo {author}
  {\bibfnamefont {L.}~\bibnamefont {Tancredi}},\ }\href {\doibase
  10.1103/PhysRevLett.113.212001} {\bibfield  {journal} {\bibinfo  {journal}
  {Phys. Rev. Lett.}\ }\textbf {\bibinfo {volume} {113}},\ \bibinfo {pages}
  {212001} (\bibinfo {year} {2014})},\ \Eprint {http://arxiv.org/abs/1408.5243}
  {arXiv:1408.5243 [hep-ph]} \BibitemShut {NoStop}%
\bibitem [{\citenamefont {Grazzini}\ \emph
  {et~al.}(2016{\natexlab{b}})\citenamefont {Grazzini}, \citenamefont
  {Kallweit}, \citenamefont {Rathlev},\ and\ \citenamefont
  {Wiesemann}}]{Grazzini:2016swo}%
  \BibitemOpen
  \bibfield  {author} {\bibinfo {author} {\bibfnamefont {M.}~\bibnamefont
  {Grazzini}}, \bibinfo {author} {\bibfnamefont {S.}~\bibnamefont {Kallweit}},
  \bibinfo {author} {\bibfnamefont {D.}~\bibnamefont {Rathlev}}, \ and\
  \bibinfo {author} {\bibfnamefont {M.}~\bibnamefont {Wiesemann}},\ }\href
  {\doibase 10.1016/j.physletb.2016.08.017} {\bibfield  {journal} {\bibinfo
  {journal} {Phys. Lett.}\ }\textbf {\bibinfo {volume} {B761}},\ \bibinfo
  {pages} {179} (\bibinfo {year} {2016}{\natexlab{b}})},\ \Eprint
  {http://arxiv.org/abs/1604.08576} {arXiv:1604.08576 [hep-ph]} \BibitemShut
  {NoStop}%
\bibitem [{\citenamefont {Grazzini}\ \emph
  {et~al.}(2016{\natexlab{c}})\citenamefont {Grazzini}, \citenamefont
  {Kallweit},\ and\ \citenamefont {Rathlev}}]{Grazzini:2016hai}%
  \BibitemOpen
  \bibfield  {author} {\bibinfo {author} {\bibfnamefont {M.}~\bibnamefont
  {Grazzini}}, \bibinfo {author} {\bibfnamefont {S.}~\bibnamefont {Kallweit}},
  \ and\ \bibinfo {author} {\bibfnamefont {D.}~\bibnamefont {Rathlev}},\ }in\
  \href {http://inspirehep.net/record/1416850/files/arXiv:1601.06751.pdf}
  {\emph {\bibinfo {booktitle} {{Proceedings, 12th International Symposium on
  Radiative Corrections (Radcor 2015) and LoopFest XIV (Radiative Corrections
  for the LHC and Future Colliders): Los Angeles, CA, USA, June 15-19,
  2015}}}}\ (\bibinfo {year} {2016})\ \Eprint {http://arxiv.org/abs/1601.06751}
  {arXiv:1601.06751 [hep-ph]} \BibitemShut {NoStop}%
\bibitem [{\citenamefont {Grazzini}\ \emph
  {et~al.}(2015{\natexlab{a}})\citenamefont {Grazzini}, \citenamefont
  {Kallweit},\ and\ \citenamefont {Rathlev}}]{Grazzini:2015nwa}%
  \BibitemOpen
  \bibfield  {author} {\bibinfo {author} {\bibfnamefont {M.}~\bibnamefont
  {Grazzini}}, \bibinfo {author} {\bibfnamefont {S.}~\bibnamefont {Kallweit}},
  \ and\ \bibinfo {author} {\bibfnamefont {D.}~\bibnamefont {Rathlev}},\ }\href
  {\doibase 10.1007/JHEP07(2015)085} {\bibfield  {journal} {\bibinfo  {journal}
  {JHEP}\ }\textbf {\bibinfo {volume} {07}},\ \bibinfo {pages} {085} (\bibinfo
  {year} {2015}{\natexlab{a}})},\ \Eprint {http://arxiv.org/abs/1504.01330}
  {arXiv:1504.01330 [hep-ph]} \BibitemShut {NoStop}%
\bibitem [{\citenamefont {Grazzini}\ \emph {et~al.}(2014)\citenamefont
  {Grazzini}, \citenamefont {Kallweit}, \citenamefont {Rathlev},\ and\
  \citenamefont {Torre}}]{Grazzini:2013bna}%
  \BibitemOpen
  \bibfield  {author} {\bibinfo {author} {\bibfnamefont {M.}~\bibnamefont
  {Grazzini}}, \bibinfo {author} {\bibfnamefont {S.}~\bibnamefont {Kallweit}},
  \bibinfo {author} {\bibfnamefont {D.}~\bibnamefont {Rathlev}}, \ and\
  \bibinfo {author} {\bibfnamefont {A.}~\bibnamefont {Torre}},\ }\href
  {\doibase 10.1016/j.physletb.2014.02.037} {\bibfield  {journal} {\bibinfo
  {journal} {Phys. Lett.}\ }\textbf {\bibinfo {volume} {B731}},\ \bibinfo
  {pages} {204} (\bibinfo {year} {2014})},\ \Eprint
  {http://arxiv.org/abs/1309.7000} {arXiv:1309.7000 [hep-ph]} \BibitemShut
  {NoStop}%
\bibitem [{\citenamefont {Grazzini}\ \emph
  {et~al.}(2015{\natexlab{b}})\citenamefont {Grazzini}, \citenamefont
  {Kallweit},\ and\ \citenamefont {Rathlev}}]{Grazzini:2015hta}%
  \BibitemOpen
  \bibfield  {author} {\bibinfo {author} {\bibfnamefont {M.}~\bibnamefont
  {Grazzini}}, \bibinfo {author} {\bibfnamefont {S.}~\bibnamefont {Kallweit}},
  \ and\ \bibinfo {author} {\bibfnamefont {D.}~\bibnamefont {Rathlev}},\ }\href
  {\doibase 10.1016/j.physletb.2015.09.055} {\bibfield  {journal} {\bibinfo
  {journal} {Phys. Lett.}\ }\textbf {\bibinfo {volume} {B750}},\ \bibinfo
  {pages} {407} (\bibinfo {year} {2015}{\natexlab{b}})},\ \Eprint
  {http://arxiv.org/abs/1507.06257} {arXiv:1507.06257 [hep-ph]} \BibitemShut
  {NoStop}%
\bibitem [{\citenamefont {Cascioli}\ \emph
  {et~al.}(2014{\natexlab{a}})\citenamefont {Cascioli}, \citenamefont
  {Gehrmann}, \citenamefont {Grazzini}, \citenamefont {Kallweit}, \citenamefont
  {Maierhöfer}, \citenamefont {von Manteuffel}, \citenamefont {Pozzorini},
  \citenamefont {Rathlev}, \citenamefont {Tancredi},\ and\ \citenamefont
  {Weihs}}]{Cascioli:2014yka}%
  \BibitemOpen
  \bibfield  {author} {\bibinfo {author} {\bibfnamefont {F.}~\bibnamefont
  {Cascioli}}, \bibinfo {author} {\bibfnamefont {T.}~\bibnamefont {Gehrmann}},
  \bibinfo {author} {\bibfnamefont {M.}~\bibnamefont {Grazzini}}, \bibinfo
  {author} {\bibfnamefont {S.}~\bibnamefont {Kallweit}}, \bibinfo {author}
  {\bibfnamefont {P.}~\bibnamefont {Maierhöfer}}, \bibinfo {author}
  {\bibfnamefont {A.}~\bibnamefont {von Manteuffel}}, \bibinfo {author}
  {\bibfnamefont {S.}~\bibnamefont {Pozzorini}}, \bibinfo {author}
  {\bibfnamefont {D.}~\bibnamefont {Rathlev}}, \bibinfo {author} {\bibfnamefont
  {L.}~\bibnamefont {Tancredi}}, \ and\ \bibinfo {author} {\bibfnamefont
  {E.}~\bibnamefont {Weihs}},\ }\href {\doibase 10.1016/j.physletb.2014.06.056}
  {\bibfield  {journal} {\bibinfo  {journal} {Phys. Lett.}\ }\textbf {\bibinfo
  {volume} {B735}},\ \bibinfo {pages} {311} (\bibinfo {year}
  {2014}{\natexlab{a}})},\ \Eprint {http://arxiv.org/abs/1405.2219}
  {arXiv:1405.2219 [hep-ph]} \BibitemShut {NoStop}%
\bibitem [{\citenamefont {Catani}\ \emph {et~al.}(2012)\citenamefont {Catani},
  \citenamefont {Cieri}, \citenamefont {de~Florian}, \citenamefont {Ferrera},\
  and\ \citenamefont {Grazzini}}]{Catani:2011qz}%
  \BibitemOpen
  \bibfield  {author} {\bibinfo {author} {\bibfnamefont {S.}~\bibnamefont
  {Catani}}, \bibinfo {author} {\bibfnamefont {L.}~\bibnamefont {Cieri}},
  \bibinfo {author} {\bibfnamefont {D.}~\bibnamefont {de~Florian}}, \bibinfo
  {author} {\bibfnamefont {G.}~\bibnamefont {Ferrera}}, \ and\ \bibinfo
  {author} {\bibfnamefont {M.}~\bibnamefont {Grazzini}},\ }\href {\doibase
  10.1103/PhysRevLett.108.072001, 10.1103/PhysRevLett.117.089901} {\bibfield
  {journal} {\bibinfo  {journal} {Phys. Rev. Lett.}\ }\textbf {\bibinfo
  {volume} {108}},\ \bibinfo {pages} {072001} (\bibinfo {year} {2012})},\
  \bibinfo {note} {[Erratum: Phys. Rev. Lett.117,no.8,089901(2016)]},\ \Eprint
  {http://arxiv.org/abs/1110.2375} {arXiv:1110.2375 [hep-ph]} \BibitemShut
  {NoStop}%
\bibitem [{\citenamefont {Campbell}\ \emph
  {et~al.}(2016{\natexlab{a}})\citenamefont {Campbell}, \citenamefont {Ellis},
  \citenamefont {Li},\ and\ \citenamefont {Williams}}]{Campbell:2016yrh}%
  \BibitemOpen
  \bibfield  {author} {\bibinfo {author} {\bibfnamefont {J.~M.}\ \bibnamefont
  {Campbell}}, \bibinfo {author} {\bibfnamefont {R.~K.}\ \bibnamefont {Ellis}},
  \bibinfo {author} {\bibfnamefont {Y.}~\bibnamefont {Li}}, \ and\ \bibinfo
  {author} {\bibfnamefont {C.}~\bibnamefont {Williams}},\ }\href {\doibase
  10.1007/JHEP07(2016)148} {\bibfield  {journal} {\bibinfo  {journal} {JHEP}\
  }\textbf {\bibinfo {volume} {07}},\ \bibinfo {pages} {148} (\bibinfo {year}
  {2016}{\natexlab{a}})},\ \Eprint {http://arxiv.org/abs/1603.02663}
  {arXiv:1603.02663 [hep-ph]} \BibitemShut {NoStop}%
\bibitem [{\citenamefont {Grazzini}\ \emph
  {et~al.}(2015{\natexlab{c}})\citenamefont {Grazzini}, \citenamefont
  {Kallweit}, \citenamefont {Rathlev},\ and\ \citenamefont
  {Wiesemann}}]{Grazzini:2015wpa}%
  \BibitemOpen
  \bibfield  {author} {\bibinfo {author} {\bibfnamefont {M.}~\bibnamefont
  {Grazzini}}, \bibinfo {author} {\bibfnamefont {S.}~\bibnamefont {Kallweit}},
  \bibinfo {author} {\bibfnamefont {D.}~\bibnamefont {Rathlev}}, \ and\
  \bibinfo {author} {\bibfnamefont {M.}~\bibnamefont {Wiesemann}},\ }\href
  {\doibase 10.1007/JHEP08(2015)154} {\bibfield  {journal} {\bibinfo  {journal}
  {JHEP}\ }\textbf {\bibinfo {volume} {08}},\ \bibinfo {pages} {154} (\bibinfo
  {year} {2015}{\natexlab{c}})},\ \Eprint {http://arxiv.org/abs/1507.02565}
  {arXiv:1507.02565 [hep-ph]} \BibitemShut {NoStop}%
\bibitem [{\citenamefont {Dawson}\ \emph {et~al.}(2016)\citenamefont {Dawson},
  \citenamefont {Jaiswal}, \citenamefont {Li}, \citenamefont {Ramani},\ and\
  \citenamefont {Zeng}}]{Dawson:2016ysj}%
  \BibitemOpen
  \bibfield  {author} {\bibinfo {author} {\bibfnamefont {S.}~\bibnamefont
  {Dawson}}, \bibinfo {author} {\bibfnamefont {P.}~\bibnamefont {Jaiswal}},
  \bibinfo {author} {\bibfnamefont {Y.}~\bibnamefont {Li}}, \bibinfo {author}
  {\bibfnamefont {H.}~\bibnamefont {Ramani}}, \ and\ \bibinfo {author}
  {\bibfnamefont {M.}~\bibnamefont {Zeng}},\ }\href@noop {} {\bibfield
  {journal} {\bibinfo  {journal} {Submitted to: Phys. Rev. D}\ } (\bibinfo
  {year} {2016})},\ \Eprint {http://arxiv.org/abs/1606.01034} {arXiv:1606.01034
  [hep-ph]} \BibitemShut {NoStop}%
\bibitem [{\citenamefont {Binoth}\ \emph {et~al.}(2005)\citenamefont {Binoth},
  \citenamefont {Ciccolini}, \citenamefont {Kauer},\ and\ \citenamefont
  {Kramer}}]{Binoth:2005ua}%
  \BibitemOpen
  \bibfield  {author} {\bibinfo {author} {\bibfnamefont {T.}~\bibnamefont
  {Binoth}}, \bibinfo {author} {\bibfnamefont {M.}~\bibnamefont {Ciccolini}},
  \bibinfo {author} {\bibfnamefont {N.}~\bibnamefont {Kauer}}, \ and\ \bibinfo
  {author} {\bibfnamefont {M.}~\bibnamefont {Kramer}},\ }\href {\doibase
  10.1088/1126-6708/2005/03/065} {\bibfield  {journal} {\bibinfo  {journal}
  {JHEP}\ }\textbf {\bibinfo {volume} {03}},\ \bibinfo {pages} {065} (\bibinfo
  {year} {2005})},\ \Eprint {http://arxiv.org/abs/hep-ph/0503094}
  {arXiv:hep-ph/0503094 [hep-ph]} \BibitemShut {NoStop}%
\bibitem [{\citenamefont {Binoth}\ \emph {et~al.}(2006)\citenamefont {Binoth},
  \citenamefont {Ciccolini}, \citenamefont {Kauer},\ and\ \citenamefont
  {Kramer}}]{Binoth:2006mf}%
  \BibitemOpen
  \bibfield  {author} {\bibinfo {author} {\bibfnamefont {T.}~\bibnamefont
  {Binoth}}, \bibinfo {author} {\bibfnamefont {M.}~\bibnamefont {Ciccolini}},
  \bibinfo {author} {\bibfnamefont {N.}~\bibnamefont {Kauer}}, \ and\ \bibinfo
  {author} {\bibfnamefont {M.}~\bibnamefont {Kramer}},\ }\href {\doibase
  10.1088/1126-6708/2006/12/046} {\bibfield  {journal} {\bibinfo  {journal}
  {JHEP}\ }\textbf {\bibinfo {volume} {12}},\ \bibinfo {pages} {046} (\bibinfo
  {year} {2006})},\ \Eprint {http://arxiv.org/abs/hep-ph/0611170}
  {arXiv:hep-ph/0611170 [hep-ph]} \BibitemShut {NoStop}%
\bibitem [{\citenamefont {Binoth}\ \emph {et~al.}(2008)\citenamefont {Binoth},
  \citenamefont {Kauer},\ and\ \citenamefont {Mertsch}}]{Binoth:2008pr}%
  \BibitemOpen
  \bibfield  {author} {\bibinfo {author} {\bibfnamefont {T.}~\bibnamefont
  {Binoth}}, \bibinfo {author} {\bibfnamefont {N.}~\bibnamefont {Kauer}}, \
  and\ \bibinfo {author} {\bibfnamefont {P.}~\bibnamefont {Mertsch}},\ }in\
  \href {\doibase 10.3360/dis.2008.142} {\emph {\bibinfo {booktitle}
  {{Proceedings, 16th International Workshop on Deep Inelastic Scattering and
  Related Subjects (DIS 2008): London, UK, April 7-11, 2008}}}}\ (\bibinfo
  {year} {2008})\ p.\ \bibinfo {pages} {142},\ \Eprint
  {http://arxiv.org/abs/0807.0024} {arXiv:0807.0024 [hep-ph]} \BibitemShut
  {NoStop}%
\bibitem [{\citenamefont {Caola}\ \emph
  {et~al.}(2015{\natexlab{a}})\citenamefont {Caola}, \citenamefont {Melnikov},
  \citenamefont {Röntsch},\ and\ \citenamefont {Tancredi}}]{Caola:2015psa}%
  \BibitemOpen
  \bibfield  {author} {\bibinfo {author} {\bibfnamefont {F.}~\bibnamefont
  {Caola}}, \bibinfo {author} {\bibfnamefont {K.}~\bibnamefont {Melnikov}},
  \bibinfo {author} {\bibfnamefont {R.}~\bibnamefont {Röntsch}}, \ and\
  \bibinfo {author} {\bibfnamefont {L.}~\bibnamefont {Tancredi}},\ }\href
  {\doibase 10.1103/PhysRevD.92.094028} {\bibfield  {journal} {\bibinfo
  {journal} {Phys. Rev.}\ }\textbf {\bibinfo {volume} {D92}},\ \bibinfo {pages}
  {094028} (\bibinfo {year} {2015}{\natexlab{a}})},\ \Eprint
  {http://arxiv.org/abs/1509.06734} {arXiv:1509.06734 [hep-ph]} \BibitemShut
  {NoStop}%
\bibitem [{\citenamefont {Caola}\ \emph
  {et~al.}(2016{\natexlab{a}})\citenamefont {Caola}, \citenamefont {Melnikov},
  \citenamefont {Röntsch},\ and\ \citenamefont {Tancredi}}]{Caola:2015rqy}%
  \BibitemOpen
  \bibfield  {author} {\bibinfo {author} {\bibfnamefont {F.}~\bibnamefont
  {Caola}}, \bibinfo {author} {\bibfnamefont {K.}~\bibnamefont {Melnikov}},
  \bibinfo {author} {\bibfnamefont {R.}~\bibnamefont {Röntsch}}, \ and\
  \bibinfo {author} {\bibfnamefont {L.}~\bibnamefont {Tancredi}},\ }\href
  {\doibase 10.1016/j.physletb.2016.01.046} {\bibfield  {journal} {\bibinfo
  {journal} {Phys. Lett.}\ }\textbf {\bibinfo {volume} {B754}},\ \bibinfo
  {pages} {275} (\bibinfo {year} {2016}{\natexlab{a}})},\ \Eprint
  {http://arxiv.org/abs/1511.08617} {arXiv:1511.08617 [hep-ph]} \BibitemShut
  {NoStop}%
\bibitem [{\citenamefont {Caola}\ \emph
  {et~al.}(2016{\natexlab{b}})\citenamefont {Caola}, \citenamefont {Dowling},
  \citenamefont {Melnikov}, \citenamefont {Röntsch},\ and\ \citenamefont
  {Tancredi}}]{Caola:2016trd}%
  \BibitemOpen
  \bibfield  {author} {\bibinfo {author} {\bibfnamefont {F.}~\bibnamefont
  {Caola}}, \bibinfo {author} {\bibfnamefont {M.}~\bibnamefont {Dowling}},
  \bibinfo {author} {\bibfnamefont {K.}~\bibnamefont {Melnikov}}, \bibinfo
  {author} {\bibfnamefont {R.}~\bibnamefont {Röntsch}}, \ and\ \bibinfo
  {author} {\bibfnamefont {L.}~\bibnamefont {Tancredi}},\ }\href@noop {} {\
  (\bibinfo {year} {2016}{\natexlab{b}})},\ \Eprint
  {http://arxiv.org/abs/1605.04610} {arXiv:1605.04610 [hep-ph]} \BibitemShut
  {NoStop}%
\bibitem [{\citenamefont {Campbell}\ \emph
  {et~al.}(2016{\natexlab{b}})\citenamefont {Campbell}, \citenamefont {Ellis},
  \citenamefont {Czakon},\ and\ \citenamefont {Kirchner}}]{Campbell:2016ivq}%
  \BibitemOpen
  \bibfield  {author} {\bibinfo {author} {\bibfnamefont {J.~M.}\ \bibnamefont
  {Campbell}}, \bibinfo {author} {\bibfnamefont {R.~K.}\ \bibnamefont {Ellis}},
  \bibinfo {author} {\bibfnamefont {M.}~\bibnamefont {Czakon}}, \ and\ \bibinfo
  {author} {\bibfnamefont {S.}~\bibnamefont {Kirchner}},\ }\href@noop {} {\
  (\bibinfo {year} {2016}{\natexlab{b}})},\ \Eprint
  {http://arxiv.org/abs/1605.01380} {arXiv:1605.01380 [hep-ph]} \BibitemShut
  {NoStop}%
\bibitem [{\citenamefont {Kauer}\ and\ \citenamefont
  {Passarino}(2012)}]{Kauer:2012hd}%
  \BibitemOpen
  \bibfield  {author} {\bibinfo {author} {\bibfnamefont {N.}~\bibnamefont
  {Kauer}}\ and\ \bibinfo {author} {\bibfnamefont {G.}~\bibnamefont
  {Passarino}},\ }\href {\doibase 10.1007/JHEP08(2012)116} {\bibfield
  {journal} {\bibinfo  {journal} {JHEP}\ }\textbf {\bibinfo {volume} {08}},\
  \bibinfo {pages} {116} (\bibinfo {year} {2012})},\ \Eprint
  {http://arxiv.org/abs/1206.4803} {arXiv:1206.4803 [hep-ph]} \BibitemShut
  {NoStop}%
\bibitem [{\citenamefont {\texttt{http://mcfm.fnal.gov}}()}]{MCFM}%
  \BibitemOpen
  \bibfield  {author} {\bibinfo {author} {\bibnamefont
  {\texttt{http://mcfm.fnal.gov}}},\ }\href@noop {} {\ }\BibitemShut {NoStop}%
\bibitem [{\citenamefont {Caola}\ \emph
  {et~al.}(2015{\natexlab{b}})\citenamefont {Caola}, \citenamefont {Henn},
  \citenamefont {Melnikov}, \citenamefont {Smirnov},\ and\ \citenamefont
  {Smirnov}}]{Caola:2015ila}%
  \BibitemOpen
  \bibfield  {author} {\bibinfo {author} {\bibfnamefont {F.}~\bibnamefont
  {Caola}}, \bibinfo {author} {\bibfnamefont {J.~M.}\ \bibnamefont {Henn}},
  \bibinfo {author} {\bibfnamefont {K.}~\bibnamefont {Melnikov}}, \bibinfo
  {author} {\bibfnamefont {A.~V.}\ \bibnamefont {Smirnov}}, \ and\ \bibinfo
  {author} {\bibfnamefont {V.~A.}\ \bibnamefont {Smirnov}},\ }\href {\doibase
  10.1007/JHEP06(2015)129} {\bibfield  {journal} {\bibinfo  {journal} {JHEP}\
  }\textbf {\bibinfo {volume} {06}},\ \bibinfo {pages} {129} (\bibinfo {year}
  {2015}{\natexlab{b}})},\ \Eprint {http://arxiv.org/abs/1503.08759}
  {arXiv:1503.08759 [hep-ph]} \BibitemShut {NoStop}%
\bibitem [{\citenamefont {von Manteuffel}\ and\ \citenamefont
  {Tancredi}(2015)}]{vonManteuffel:2015msa}%
  \BibitemOpen
  \bibfield  {author} {\bibinfo {author} {\bibfnamefont {A.}~\bibnamefont {von
  Manteuffel}}\ and\ \bibinfo {author} {\bibfnamefont {L.}~\bibnamefont
  {Tancredi}},\ }\href {\doibase 10.1007/JHEP06(2015)197} {\bibfield  {journal}
  {\bibinfo  {journal} {JHEP}\ }\textbf {\bibinfo {volume} {06}},\ \bibinfo
  {pages} {197} (\bibinfo {year} {2015})},\ \Eprint
  {http://arxiv.org/abs/1503.08835} {arXiv:1503.08835 [hep-ph]} \BibitemShut
  {NoStop}%
\bibitem [{\citenamefont {\texttt{https://vvamp.hepforge.org/}}()}]{ggvvamp}%
  \BibitemOpen
  \bibfield  {author} {\bibinfo {author} {\bibnamefont
  {\texttt{https://vvamp.hepforge.org/}}},\ }\href@noop {} {\ }\BibitemShut
  {NoStop}%
\bibitem [{\citenamefont {Cascioli}\ \emph
  {et~al.}(2014{\natexlab{b}})\citenamefont {Cascioli}, \citenamefont {Höche},
  \citenamefont {Krauss}, \citenamefont {Maierhöfer}, \citenamefont
  {Pozzorini},\ and\ \citenamefont {Siegert}}]{Cascioli:2013gfa}%
  \BibitemOpen
  \bibfield  {author} {\bibinfo {author} {\bibfnamefont {F.}~\bibnamefont
  {Cascioli}}, \bibinfo {author} {\bibfnamefont {S.}~\bibnamefont {Höche}},
  \bibinfo {author} {\bibfnamefont {F.}~\bibnamefont {Krauss}}, \bibinfo
  {author} {\bibfnamefont {P.}~\bibnamefont {Maierhöfer}}, \bibinfo {author}
  {\bibfnamefont {S.}~\bibnamefont {Pozzorini}}, \ and\ \bibinfo {author}
  {\bibfnamefont {F.}~\bibnamefont {Siegert}},\ }\href {\doibase
  10.1007/JHEP01(2014)046} {\bibfield  {journal} {\bibinfo  {journal} {JHEP}\
  }\textbf {\bibinfo {volume} {01}},\ \bibinfo {pages} {046} (\bibinfo {year}
  {2014}{\natexlab{b}})},\ \Eprint {http://arxiv.org/abs/1309.0500}
  {arXiv:1309.0500 [hep-ph]} \BibitemShut {NoStop}%
\bibitem [{\citenamefont {Alioli}\ \emph {et~al.}(2010)\citenamefont {Alioli},
  \citenamefont {Nason}, \citenamefont {Oleari},\ and\ \citenamefont
  {Re}}]{Alioli:2010xd}%
  \BibitemOpen
  \bibfield  {author} {\bibinfo {author} {\bibfnamefont {S.}~\bibnamefont
  {Alioli}}, \bibinfo {author} {\bibfnamefont {P.}~\bibnamefont {Nason}},
  \bibinfo {author} {\bibfnamefont {C.}~\bibnamefont {Oleari}}, \ and\ \bibinfo
  {author} {\bibfnamefont {E.}~\bibnamefont {Re}},\ }\href {\doibase
  10.1007/JHEP06(2010)043} {\bibfield  {journal} {\bibinfo  {journal} {JHEP}\
  }\textbf {\bibinfo {volume} {06}},\ \bibinfo {pages} {043} (\bibinfo {year}
  {2010})},\ \Eprint {http://arxiv.org/abs/1002.2581} {arXiv:1002.2581
  [hep-ph]} \BibitemShut {NoStop}%
\bibitem [{\citenamefont {Cullen}\ \emph {et~al.}(2012)\citenamefont {Cullen},
  \citenamefont {Greiner}, \citenamefont {Heinrich}, \citenamefont {Luisoni},
  \citenamefont {Mastrolia} \emph {et~al.}}]{Cullen:2011ac}%
  \BibitemOpen
  \bibfield  {author} {\bibinfo {author} {\bibfnamefont {G.}~\bibnamefont
  {Cullen}}, \bibinfo {author} {\bibfnamefont {N.}~\bibnamefont {Greiner}},
  \bibinfo {author} {\bibfnamefont {G.}~\bibnamefont {Heinrich}}, \bibinfo
  {author} {\bibfnamefont {G.}~\bibnamefont {Luisoni}}, \bibinfo {author}
  {\bibfnamefont {P.}~\bibnamefont {Mastrolia}},  \emph {et~al.},\ }\href
  {\doibase 10.1140/epjc/s10052-012-1889-1} {\bibfield  {journal} {\bibinfo
  {journal} {Eur.Phys.J.}\ }\textbf {\bibinfo {volume} {C72}},\ \bibinfo
  {pages} {1889} (\bibinfo {year} {2012})},\ \Eprint
  {http://arxiv.org/abs/1111.2034} {arXiv:1111.2034 [hep-ph]} \BibitemShut
  {NoStop}%
\bibitem [{\citenamefont {Cullen}\ \emph {et~al.}(2014)\citenamefont {Cullen},
  \citenamefont {van Deurzen}, \citenamefont {Greiner}, \citenamefont
  {Heinrich}, \citenamefont {Luisoni} \emph {et~al.}}]{Cullen:2014yla}%
  \BibitemOpen
  \bibfield  {author} {\bibinfo {author} {\bibfnamefont {G.}~\bibnamefont
  {Cullen}}, \bibinfo {author} {\bibfnamefont {H.}~\bibnamefont {van Deurzen}},
  \bibinfo {author} {\bibfnamefont {N.}~\bibnamefont {Greiner}}, \bibinfo
  {author} {\bibfnamefont {G.}~\bibnamefont {Heinrich}}, \bibinfo {author}
  {\bibfnamefont {G.}~\bibnamefont {Luisoni}},  \emph {et~al.},\ }\href
  {\doibase 10.1140/epjc/s10052-014-3001-5} {\bibfield  {journal} {\bibinfo
  {journal} {Eur.Phys.J.}\ }\textbf {\bibinfo {volume} {C74}},\ \bibinfo
  {pages} {3001} (\bibinfo {year} {2014})},\ \Eprint
  {http://arxiv.org/abs/1404.7096} {arXiv:1404.7096 [hep-ph]} \BibitemShut
  {NoStop}%
\bibitem [{\citenamefont {van Deurzen}\ \emph {et~al.}(2014)\citenamefont {van
  Deurzen}, \citenamefont {Luisoni}, \citenamefont {Mastrolia}, \citenamefont
  {Mirabella}, \citenamefont {Ossola} \emph {et~al.}}]{vanDeurzen:2013saa}%
  \BibitemOpen
  \bibfield  {author} {\bibinfo {author} {\bibfnamefont {H.}~\bibnamefont {van
  Deurzen}}, \bibinfo {author} {\bibfnamefont {G.}~\bibnamefont {Luisoni}},
  \bibinfo {author} {\bibfnamefont {P.}~\bibnamefont {Mastrolia}}, \bibinfo
  {author} {\bibfnamefont {E.}~\bibnamefont {Mirabella}}, \bibinfo {author}
  {\bibfnamefont {G.}~\bibnamefont {Ossola}},  \emph {et~al.},\ }\href
  {\doibase 10.1007/JHEP03(2014)115} {\bibfield  {journal} {\bibinfo  {journal}
  {JHEP}\ }\textbf {\bibinfo {volume} {1403}},\ \bibinfo {pages} {115}
  (\bibinfo {year} {2014})},\ \Eprint {http://arxiv.org/abs/1312.6678}
  {arXiv:1312.6678 [hep-ph]} \BibitemShut {NoStop}%
\bibitem [{\citenamefont {Peraro}(2014)}]{Peraro:2014cba}%
  \BibitemOpen
  \bibfield  {author} {\bibinfo {author} {\bibfnamefont {T.}~\bibnamefont
  {Peraro}},\ }\href {\doibase 10.1016/j.cpc.2014.06.017} {\bibfield  {journal}
  {\bibinfo  {journal} {Comput.Phys.Commun.}\ }\textbf {\bibinfo {volume}
  {185}},\ \bibinfo {pages} {2771} (\bibinfo {year} {2014})},\ \Eprint
  {http://arxiv.org/abs/1403.1229} {arXiv:1403.1229 [hep-ph]} \BibitemShut
  {NoStop}%
\bibitem [{\citenamefont {Cascioli}\ \emph {et~al.}(2012)\citenamefont
  {Cascioli}, \citenamefont {Maierhofer},\ and\ \citenamefont
  {Pozzorini}}]{Cascioli:2011va}%
  \BibitemOpen
  \bibfield  {author} {\bibinfo {author} {\bibfnamefont {F.}~\bibnamefont
  {Cascioli}}, \bibinfo {author} {\bibfnamefont {P.}~\bibnamefont
  {Maierhofer}}, \ and\ \bibinfo {author} {\bibfnamefont {S.}~\bibnamefont
  {Pozzorini}},\ }\href {\doibase 10.1103/PhysRevLett.108.111601} {\bibfield
  {journal} {\bibinfo  {journal} {Phys. Rev. Lett.}\ }\textbf {\bibinfo
  {volume} {108}},\ \bibinfo {pages} {111601} (\bibinfo {year} {2012})},\
  \Eprint {http://arxiv.org/abs/1111.5206} {arXiv:1111.5206 [hep-ph]}
  \BibitemShut {NoStop}%
\bibitem [{\citenamefont {Sjostrand}\ \emph {et~al.}(2008)\citenamefont
  {Sjostrand}, \citenamefont {Mrenna},\ and\ \citenamefont
  {Skands}}]{Sjostrand:2007gs}%
  \BibitemOpen
  \bibfield  {author} {\bibinfo {author} {\bibfnamefont {T.}~\bibnamefont
  {Sjostrand}}, \bibinfo {author} {\bibfnamefont {S.}~\bibnamefont {Mrenna}}, \
  and\ \bibinfo {author} {\bibfnamefont {P.~Z.}\ \bibnamefont {Skands}},\
  }\href {\doibase 10.1016/j.cpc.2008.01.036} {\bibfield  {journal} {\bibinfo
  {journal} {Comput.Phys.Commun.}\ }\textbf {\bibinfo {volume} {178}},\
  \bibinfo {pages} {852} (\bibinfo {year} {2008})},\ \Eprint
  {http://arxiv.org/abs/0710.3820} {arXiv:0710.3820 [hep-ph]} \BibitemShut
  {NoStop}%
\bibitem [{\citenamefont {Ball}\ \emph {et~al.}(2015)\citenamefont {Ball} \emph
  {et~al.}}]{Ball:2014uwa}%
  \BibitemOpen
  \bibfield  {author} {\bibinfo {author} {\bibfnamefont {R.~D.}\ \bibnamefont
  {Ball}} \emph {et~al.} (\bibinfo {collaboration} {NNPDF}),\ }\href {\doibase
  10.1007/JHEP04(2015)040} {\bibfield  {journal} {\bibinfo  {journal} {JHEP}\
  }\textbf {\bibinfo {volume} {04}},\ \bibinfo {pages} {040} (\bibinfo {year}
  {2015})},\ \Eprint {http://arxiv.org/abs/1410.8849} {arXiv:1410.8849
  [hep-ph]} \BibitemShut {NoStop}%
\bibitem [{\citenamefont {Cacciari}\ \emph {et~al.}(2008)\citenamefont
  {Cacciari}, \citenamefont {Salam},\ and\ \citenamefont
  {Soyez}}]{Cacciari:2008gp}%
  \BibitemOpen
  \bibfield  {author} {\bibinfo {author} {\bibfnamefont {M.}~\bibnamefont
  {Cacciari}}, \bibinfo {author} {\bibfnamefont {G.~P.}\ \bibnamefont {Salam}},
  \ and\ \bibinfo {author} {\bibfnamefont {G.}~\bibnamefont {Soyez}},\ }\href
  {\doibase 10.1088/1126-6708/2008/04/063} {\bibfield  {journal} {\bibinfo
  {journal} {JHEP}\ }\textbf {\bibinfo {volume} {04}},\ \bibinfo {pages} {063}
  (\bibinfo {year} {2008})},\ \Eprint {http://arxiv.org/abs/0802.1189}
  {arXiv:0802.1189 [hep-ph]} \BibitemShut {NoStop}%
\bibitem [{\citenamefont {Cacciari}\ and\ \citenamefont
  {Salam}(2006)}]{Cacciari:2005hq}%
  \BibitemOpen
  \bibfield  {author} {\bibinfo {author} {\bibfnamefont {M.}~\bibnamefont
  {Cacciari}}\ and\ \bibinfo {author} {\bibfnamefont {G.~P.}\ \bibnamefont
  {Salam}},\ }\href {\doibase 10.1016/j.physletb.2006.08.037} {\bibfield
  {journal} {\bibinfo  {journal} {Phys.Lett.}\ }\textbf {\bibinfo {volume}
  {B641}},\ \bibinfo {pages} {57} (\bibinfo {year} {2006})},\ \Eprint
  {http://arxiv.org/abs/hep-ph/0512210} {arXiv:hep-ph/0512210 [hep-ph]}
  \BibitemShut {NoStop}%
\bibitem [{\citenamefont {Cacciari}\ \emph {et~al.}(2012)\citenamefont
  {Cacciari}, \citenamefont {Salam},\ and\ \citenamefont
  {Soyez}}]{Cacciari:2011ma}%
  \BibitemOpen
  \bibfield  {author} {\bibinfo {author} {\bibfnamefont {M.}~\bibnamefont
  {Cacciari}}, \bibinfo {author} {\bibfnamefont {G.~P.}\ \bibnamefont {Salam}},
  \ and\ \bibinfo {author} {\bibfnamefont {G.}~\bibnamefont {Soyez}},\ }\href
  {\doibase 10.1140/epjc/s10052-012-1896-2} {\bibfield  {journal} {\bibinfo
  {journal} {Eur.Phys.J.}\ }\textbf {\bibinfo {volume} {C72}},\ \bibinfo
  {pages} {1896} (\bibinfo {year} {2012})},\ \Eprint
  {http://arxiv.org/abs/1111.6097} {arXiv:1111.6097 [hep-ph]} \BibitemShut
  {NoStop}%
\bibitem [{\citenamefont {Bolognesi}\ \emph {et~al.}(2012)\citenamefont
  {Bolognesi}, \citenamefont {Gao}, \citenamefont {Gritsan}, \citenamefont
  {Melnikov}, \citenamefont {Schulze}, \citenamefont {Tran},\ and\
  \citenamefont {Whitbeck}}]{Bolognesi:2012mm}%
  \BibitemOpen
  \bibfield  {author} {\bibinfo {author} {\bibfnamefont {S.}~\bibnamefont
  {Bolognesi}}, \bibinfo {author} {\bibfnamefont {Y.}~\bibnamefont {Gao}},
  \bibinfo {author} {\bibfnamefont {A.~V.}\ \bibnamefont {Gritsan}}, \bibinfo
  {author} {\bibfnamefont {K.}~\bibnamefont {Melnikov}}, \bibinfo {author}
  {\bibfnamefont {M.}~\bibnamefont {Schulze}}, \bibinfo {author} {\bibfnamefont
  {N.~V.}\ \bibnamefont {Tran}}, \ and\ \bibinfo {author} {\bibfnamefont
  {A.}~\bibnamefont {Whitbeck}},\ }\href {\doibase 10.1103/PhysRevD.86.095031}
  {\bibfield  {journal} {\bibinfo  {journal} {Phys. Rev.}\ }\textbf {\bibinfo
  {volume} {D86}},\ \bibinfo {pages} {095031} (\bibinfo {year} {2012})},\
  \Eprint {http://arxiv.org/abs/1208.4018} {arXiv:1208.4018 [hep-ph]}
  \BibitemShut {NoStop}%
\bibitem [{\citenamefont {Alioli}\ \emph
  {et~al.}(2009{\natexlab{a}})\citenamefont {Alioli}, \citenamefont {Nason},
  \citenamefont {Oleari},\ and\ \citenamefont {Re}}]{Alioli:2008tz}%
  \BibitemOpen
  \bibfield  {author} {\bibinfo {author} {\bibfnamefont {S.}~\bibnamefont
  {Alioli}}, \bibinfo {author} {\bibfnamefont {P.}~\bibnamefont {Nason}},
  \bibinfo {author} {\bibfnamefont {C.}~\bibnamefont {Oleari}}, \ and\ \bibinfo
  {author} {\bibfnamefont {E.}~\bibnamefont {Re}},\ }\href {\doibase
  10.1088/1126-6708/2009/04/002} {\bibfield  {journal} {\bibinfo  {journal}
  {JHEP}\ }\textbf {\bibinfo {volume} {0904}},\ \bibinfo {pages} {002}
  (\bibinfo {year} {2009}{\natexlab{a}})},\ \Eprint
  {http://arxiv.org/abs/0812.0578} {arXiv:0812.0578 [hep-ph]} \BibitemShut
  {NoStop}%
\bibitem [{\citenamefont {Alioli}\ \emph
  {et~al.}(2009{\natexlab{b}})\citenamefont {Alioli}, \citenamefont {Nason},
  \citenamefont {Oleari},\ and\ \citenamefont {Re}}]{Alioli:2009je}%
  \BibitemOpen
  \bibfield  {author} {\bibinfo {author} {\bibfnamefont {S.}~\bibnamefont
  {Alioli}}, \bibinfo {author} {\bibfnamefont {P.}~\bibnamefont {Nason}},
  \bibinfo {author} {\bibfnamefont {C.}~\bibnamefont {Oleari}}, \ and\ \bibinfo
  {author} {\bibfnamefont {E.}~\bibnamefont {Re}},\ }\href {\doibase
  10.1088/1126-6708/2009/09/111} {\bibfield  {journal} {\bibinfo  {journal}
  {JHEP}\ }\textbf {\bibinfo {volume} {09}},\ \bibinfo {pages} {111} (\bibinfo
  {year} {2009}{\natexlab{b}})},\ \Eprint {http://arxiv.org/abs/0907.4076}
  {arXiv:0907.4076 [hep-ph]} \BibitemShut {NoStop}%
\bibitem [{\citenamefont {ATLAS-CONF-2013-020}(2013)}]{ATLAS:2013gma}%
  \BibitemOpen
  \bibfield  {author} {\bibinfo {author} {\bibnamefont {ATLAS-CONF-2013-020}},\
  }\href@noop {} {\  (\bibinfo {year} {2013})}\BibitemShut {NoStop}%
\end{thebibliography}%

\end{document}